\documentclass[12pt]{article}
\usepackage{amsmath}
\usepackage{amssymb}
\usepackage{amstext}
\usepackage{amsfonts}
\usepackage{graphicx,epsfig}
\usepackage{indentfirst}

\newtheorem{theorem}{Theorem}
\begin{document}

\title{Classification of the Entangled States of $2 \times N \times N$ }
\author{Shuo Cheng$^{a}$\footnote{chengshuo05@mails.gucas.ac.cn},\;
Junli Li$^{a}$\footnote{jlli04@mails.gucas.ac.cn}\; ~and
Cong-Feng Qiao$^{a,b}$\footnote{qiaocf@gucas.ac.cn, corresponding author}\\[0.5cm]
{\small $a)$ Dept. of Physics, Graduate
University, the Chinese Academy of Sciences}  \\
{\small YuQuan Road 19A, 100049, Beijing, China}\\
{\small $b)$ Theoretical Physics Center for Science Facilities
(TPCSF), CAS}\\
}
\date{}
\maketitle

\begin{abstract}
We develop a novel method in classifying the multipartite
entanglement state of $2\times N\times N$ under stochastic local
operation and classical communication. In this method, all
inequivalent classes of true entangled state can be assorted
directly without knowing the classification information of lower
dimension ones for any given dimension $N$. It also gives a nature
explanation for the non-local parameters remaining in the
entanglement classes while $N\geq 4$\ .
\end{abstract}

\section{Introduction}

Entanglement is at the heart of the quantum information theory (QIT)
and is now thought as a physical resource to realize quantum
information tasks, such as quantum cryptography
\cite{cryp-1,cryp-2}, superdense coding
\cite{superdense-1,superdense-2}, and quantum computation
\cite{quantum-computation}, etc. Moreover, the study of entanglement
may also improve our knowledge about quantum non-locality
\cite{ghz}. Among it the investigation on the classification of
multipartite entanglement is of particular interest in QIT.
According to QIT, two quantum states can be employed to carry on the
same task while they are thought to be equivalent in the meaning of
mutually convertible under Stochastic Local Operations and Classical
Communication (SLOCC) \cite{three-qubit}.

Nevertheless, in practice the classification of multipartite
entanglement in high dimension in the Hilbert space is generally
mathematically difficult \cite{four-qubit}. It was found that the
matrix decomposition method \cite{Jordan-c,Pic-three-qubit} keeps to
be a useful tool as in the two-partite case. A widely adopted
philosophy in dealing with this issue is first to classify the state
in lower dimension (or less partite) and then extend to the higher
dimension \cite{range-1,range-2} (or more partite
\cite{inductive-3,inductive-4}) cases in an inductive way. However,
nontrivial aspect emerges as the dimension increases, i.e. some
non-local parameters may nest in the entangled states
\cite{non-local,non-local1}. In recent years, investigations on the
classification of $2\times M\times N$ states were performed
\cite{range-1,range-2}, where $M$ and $N$ are dimensions of two
partites in three-partite entangled states. Based on the ``range
criterion", an iterated method was introduced to determine all
classes of true entangled states of the $2\times M\times N$ system
in Refs.\cite{range-1,range-2}. In this scenario the entanglement
classes of high dimensional states can be obtained through the low
dimensional ones. That is, first generate all the possible
entanglement classes under invertible local operator (ILO) by the
classification information of lower dimensional ones, then use the
``range criterion" to find out the inequivalent classes of true
entanglement among all the possible entangled classes, which tends
to be a formidable task with the increase of dimensions. The main
trait of this scenario is that the lower dimensional entanglement
classes are prerequisite for the follow-up classification. As
mentioned in Ref.\cite{range-2} the classification of entangled
state of $2\times M\times N$ becomes more subtle when $M = N$. In
this case the permutations of the two N-dimensional partites may be
assorted into different classes.

In this work, we present a straightforward method in fully
classifying the entanglement states in $2\times N\times N$
configuration. The asymmetry of the two $N$-dimensional partites
shows up in one of the classes. We develop a cubic grid form for the
quantum state, in which the entangled classes that have continuous
parameters can be explained naturally. This gives an instructive
insight on the entanglement classes of 4 or more partites which also
have non-local parameters \cite{three-qubit,inductive-4}.

The paper is arranged as follows: after the introduction section, we
represent the entangled state in a general form in section 2, by
which the true entangled state of $2\times N \times N$ can be
expressed in a matrix pair. With the definitions given in section 2,
the true entangled state of $2\times N \times N$ can be fully
classified according to the theorems given in section 3. In section
4, two examples on how to employ the novel classification method are
presented. We show, in a typical case of $2\times 5 \times 5$, that
the non-local parameters generally may exist in high dimensional or
multi-partite entangled state. The last section is remained for a
brief summary.

\section{Representation of the entangled states of $2\times N\times N$}

An arbitrary two-partite state in dimension of $M$ times $N$ can be
expressed in the following form
\begin{eqnarray}
|\Psi_{M\times N}\rangle & = & \gamma_{11}|11\rangle +
\gamma_{12}|12\rangle + \cdots \gamma_{1N}|1N \rangle + \nonumber \\
&  & \gamma_{21}|21\rangle +
\gamma_{22}|22\rangle + \cdots +\gamma_{2N}|2N\rangle + \nonumber \\
& & \;\; \vdots \nonumber \\ & & \gamma_{M1}|M1\rangle +
\gamma_{M2}|M2\rangle + \cdots \gamma_{MN}|MN \rangle \; ,
\label{general-state}
\end{eqnarray}
where $\gamma_{ij} \in \mathbb{C}$, are a series of complex numbers.
Eq.(\ref{general-state}) can be further expressed in a more compact
form
\begin{eqnarray}
|\Psi_{M\times N}\rangle & = & (|\;\!1\rangle, |\,2\rangle, \cdots ,
|\,M\rangle) \left(
  \begin{array}{cccc}
    \gamma_{11} & \gamma_{12} & \cdots & \gamma_{1N} \\
    \gamma_{21} & \gamma_{22} & \cdots & \gamma_{2N} \\
    \vdots & \vdots & \ddots & \vdots \\
    \gamma_{M1} & \gamma_{M2} & \cdots & \gamma_{MN} \\
  \end{array}
\right)
\left(
\begin{array}{c}
|\;\!1\rangle \\
|\,2\rangle \\
\vdots \\
|\,N\rangle \\
\end{array}
\right) \nonumber \\
 & \equiv & \psi_1^{\rm{T}} \Gamma_{\{i,j\}} \psi_2 = Tr[\Gamma_{\{i,j\}}
 \; \psi_2 \otimes \psi_1^{\text{T}}]\; .\label{2ent-state}
\end{eqnarray}
Here, $\Gamma_{\{i,j\}}$ denotes the $M\times N$ complex matrix,
which can also be treated as a tensor of rank two, and $\otimes$ is
the symbol of direct product. Obviously, the feature of a $M\times
N$ pure state is characterized by the rank-two tensor
$\Gamma_{\{i,j\}}$. Similarly, the state of $2\times N \times N$ may
be expressed in a traced form
\begin{eqnarray}
|\Psi_{2\times N\times N}\rangle = Tr[\Gamma_{\{i,j,k\}}\; \psi_2
\otimes \psi_1^{\mathrm{T}} \otimes \psi_0^{\mathrm{T}}] \; ,
\label{P-gamma}
\end{eqnarray}
where, $\psi_0$ is a 2-dimensional vector and $\psi_{1,2}$ are
N-dimensional vectors, representing the constituent states in
Hilbert space. The $2 \times N\times N$ matrix $\Gamma_{\{i,j,k\}}$,
which can also be taken as a rank-three tensor, reads
\begin{eqnarray}
\Gamma_{\{i,j,k\}} \; = \; \left(
  \begin{array}{cccc}
    \gamma_{111} & \gamma_{112} & \cdots & \gamma_{11N} \\
    \gamma_{121} & \gamma_{122} & \cdots & \gamma_{12N} \\
    \vdots & \vdots & \ddots & \vdots \\
    \gamma_{1N1} & \gamma_{1N2} & \cdots & \gamma_{1NN} \\
    \gamma_{211} & \gamma_{212} & \cdots & \gamma_{21N} \\
    \gamma_{221} & \gamma_{222} & \cdots & \gamma_{22N} \\
    \vdots & \vdots & \ddots & \vdots \\
    \gamma_{2N1} & \gamma_{2N2} & \cdots & \gamma_{2NN} \\
  \end{array}
\right)  =  \left\lgroup
\begin{array}{c}
\Gamma_{\{1,l,m\}} \\
\Gamma_{\{2,l,m\}} \\
\end{array}
\right\rgroup
\;\equiv\; \left\lgroup
\begin{array}{c}
\Gamma_{\!1} \\
\Gamma_{\!\!2} \\
\end{array}
\right\rgroup \label{3ent-matrix} \; .
\end{eqnarray}
Here, $\Gamma_{\{1,l,m\}}$ and $\Gamma_{\{2,l,m\}}$ are in fact
tensors of rank two, which are represented by $N\times N$ complex
matrices $\Gamma_{\!1,2}$. The $\Gamma_1$ and $\Gamma_2$ stand for
the upper and lower $N$-line blocks of matrix $\Gamma_{\{i,j,k\}}$,
respectively.

From (\ref{P-gamma}) and (\ref{3ent-matrix}) we know that the
information of the state $2 \times N \times N$ is involved in the
matrix pair $(\Gamma_{\!1}\,, \Gamma_{\!\!2})$. Therefore, in the
aim of classification we can specify the typical entangled state by
a `matrix vector', that is
\begin{eqnarray}
|\Psi_{2\times N\times N}\rangle \; \doteq\; \left\lgroup
\begin{array}{c}
\Gamma_{\!1} \\
\Gamma_{\!\!2} \\
\end{array}
\right\rgroup\; ,\label{M-Vector}
\end{eqnarray}
where, the symbol ~$\doteq$~ stands for ``is represented by".

Generally speaking, two states are said to be SLOCC equivalent if
they are connected by ILOs \cite{three-qubit}. For instance, in the
case of bipartite entanglement, suppose the two partites are
transformed under two invertible local operators $P$ and $Q$, i.e. $
\psi_1'= P^{\mathrm{T}} \psi_{1}, \psi_2' = Q \psi_{2}$, then from
Eq.(\ref{2ent-state}) a SLOCC equivalent state to this bipartite
entangled state reads as
\begin{eqnarray}
|\Psi'_{M\times N}\rangle & = & \psi_1'^{\rm{T}} \Gamma_{\{i,j\}}
\psi'_2 \nonumber\\ & = & Tr[P\Gamma_{\{i,j\}}Q \;\psi_2 \otimes
\psi_1^{\rm{T}}] \nonumber \\ & = & \psi_1^{\rm{T}}
\Gamma_{\{i,j\}}' \psi_2 \; .
\end{eqnarray}
From above expression, we see that two SLOCC equivalent states are
in fact connected only by the transformation of the matrix
$\Gamma_{\{i,j\}}$ in Eq.(\ref{2ent-state}) like
\begin{eqnarray}
\Gamma'_{\{i,j\}} = P\; \Gamma_{\{i,j\}}\; Q\; .
\end{eqnarray}
Similarly, two SLOCC equivalent $2\times N\times N$ states
${|\Psi'_{2\times N\times N}\rangle}$ and $|\Psi_{2\times N\times
N}\rangle$ are also connected by the transformation of the matrix
$\Gamma_{\{i,j,k\}}$ in Eq.(\ref{3ent-matrix}), i.e.,
\begin{eqnarray}
|\Psi'_{2\times N\times N}\rangle  & = & Tr[\Gamma_{\{i,j,k\}} \;
\psi'_2 \otimes \psi_1'^{\,\rm{T}} \otimes \psi_0'^{\,\rm{T}}]
\nonumber
\\ & = & Tr[T \otimes P \Gamma_{\{i,j,k\}} Q \;
\psi_2 \otimes \psi_1^{\rm{T}} \otimes \psi_0^{\rm{T}}] \nonumber \\
& = & Tr[\Gamma'_{\{i,j,k\}} \; \psi_2 \otimes \psi_1^{\rm{T}}
\otimes \psi_0^{\rm{T}}] \; , \label{T-Psi3}
\end{eqnarray}
where
\begin{eqnarray}
\Gamma'_{\{i,j,k\}} = \left\lgroup
\begin{array}{c}
\Gamma'_{\{1,j,k\}} \\
\Gamma'_{\{2,j,k\}} \\
\end{array}
\right\rgroup = T \left\lgroup
\begin{array}{c}
P\Gamma_{\!1}Q \\
P\Gamma_{\!\!2}Q \\
\end{array}
\right\rgroup \; . \label{transfromation-1}
\end{eqnarray}
 Here, $T$ is any invertible $2\times 2$ matrix which acts on
$\psi_0$; $P$ and $Q$ are two invertible $N\times N$ matrices acting
on $\psi_{1}$ and $\psi_{2}$, respectively. The transformation of
the first partite $T$ reads as
\begin{eqnarray}
\left\lgroup
  \begin{array}{cc}
    t_{11} & t_{12} \\
    t_{21} & t_{22} \\
  \end{array}
\right\rgroup \left\lgroup
  \begin{array}{c}
    \Gamma_{\!1} \\
    \Gamma_{\!\!2} \\
  \end{array}
\right\rgroup\; .
\end{eqnarray}
For brevity, as in (\ref{M-Vector}), equation (\ref{T-Psi3}) can be
formulated as
\begin{eqnarray}
 {|\Psi'_{2\times N\times N}\rangle}  =  \left\lgroup
  \begin{array}{cc}
    t_{11} & t_{12} \\
    t_{21} & t_{22} \\
  \end{array}
\right\rgroup \left\lgroup
  \begin{array}{c}
    P \Gamma_{\!1} Q \\
    P \Gamma_{\!\!2} Q\\
  \end{array}
\right\rgroup\; , \label{T-V3}
\end{eqnarray}
where $\psi\,$s are suppressed.

\begin{figure}[t,m,u]
\centering \scalebox{1}{\includegraphics{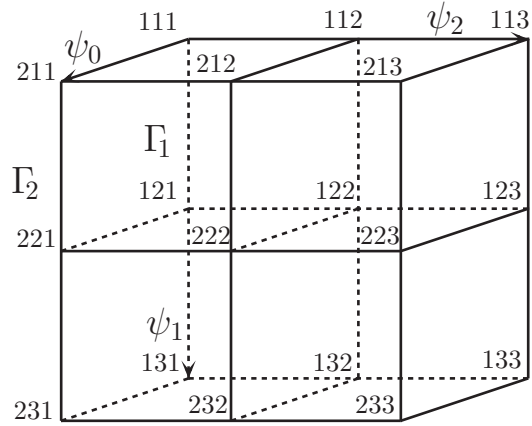}}
\caption{\small The pictorial description of $2\times 3\times 3$
state, where each node corresponds to a base vector. Assigning a
coefficient to the base vector, we then obtain the corresponding
matrix element of $\Gamma_{\!1}$ or $\Gamma_{\!\!2}$. }
\label{233-pic}
\end{figure}

To give a pictorial description of the quantum state, we take
$2\times 3\times 3$ case as an example, see Figure \ref{233-pic}.
The matrices $\Gamma_{\!1}$ and $\Gamma_{\!\!2}$ are placed in
parallel in rear and front of the cubic, respectively. Of the cubic
grid, each node corresponds to an element in the matrix pair
$(\Gamma_{\!1},\Gamma_{\!\!2})$ in Eq.(\ref{M-Vector})\ .

In matrix algebra, every ILO which acts on a given matrix can be
decomposed as a series of products of elementary operations on the
matrix, and there exist three such elementary operations
\cite{matrix-analysis}. Therefore, the matrices $T$, $P$ and $Q$ in
Eq.(\ref{T-Psi3}), which connect the two equivalent wave functions,
can be decomposed as such sequence of elementary operations. In the
pictorial language, here the three types of elementary operation
correspond to three types of manipulation of the cubic grid: {\it\bf
type 1}, interchange of two surfaces; {\it\bf type 2},
multiplication of one surface by a nonzero scalar; {\it\bf type 3},
addition of a scalar multiple of one surface to another surface.
Specifically, $T$ is responsible for the elementary operations
between front and rear; $P$ for upper and lower, $Q$ for left and
right surfaces, respectively.

According to the common definition \cite{three-qubit}, a true
$2\times N\times N$ entangled state requires the following
conditions
\begin{eqnarray}
r(\rho_{\psi_{0}})=2,\; r(\rho_{\psi_{1}}) = r(\rho_{\psi_{2}}) = N
\; , \label{rho-rank}
\end{eqnarray}
to be true, where $\rho_{i}= \mathrm{Tr}_{jk}(\rho_{ijk})$ being the
reduced density matrix. Hereafter, we denote $r$ to be the rank of
matrix. In Quantum Mechanics, to each state there corresponds a
unique state operator, the density matrix. In the representation of
matrix pair the density matrix(elements) can be expressed as
\begin{eqnarray}
\rho_{\psi_0,\psi_1,\psi_2} = \Gamma_{ijk}
\Gamma^{*}_{i{\,'}j{\,'}k{\,'}} \; ,
\end{eqnarray}
where $i,i' = 1,2; j,j' = 1,2, \cdots, N; k,k' = 1,2, \cdots, N$.
The corresponding reduced density matrix is (taking $\psi_2$ as an
example)
\begin{eqnarray}
\rho_{\psi_{2}} & = & \mathrm{Tr}_{\psi_0,\psi_1} ( \rho_{\psi_0,\psi_1,\psi_2}) \nonumber \\
& = & \sum_{ij} \Gamma_{ijk} \Gamma^{*}_{i j k'}\nonumber \\
& = & \sum_{i} (\Gamma_{i}^{\dagger}\Gamma_{i})_{k'k} \; .
\label{density-cube-nn}
\end{eqnarray}
If Det$(\rho_{\psi_2}) \neq 0$, then we know $r(\rho_{\psi_2}) = N
$.

\section{Classification of the tripartite entangled state  $2\times N\times N$}

With the above preparation, we can now proceed to classify the
$2\times N\times N$ state. Generically, the whole space of the state
$(\Gamma_{\!1},\Gamma_{\!\!2})$ can be partitioned into numbers of
inequivalent sets by different $l$ and $n$.
\begin{eqnarray}
&&(\Gamma_{\!1},\Gamma_{\!\!2})= \{C_{n,\;l}\}\; ,\\
&& C_{n,\;l}=\{ (\Gamma_{\!1},\Gamma_{\!\!2})|\;
r_{\mathrm{max}}(\alpha_1 \Gamma_{\!1} + \beta_1 \Gamma_{\!\!2}) =
n, r_{\mathrm{min}}(\alpha_2 \Gamma_{\!1} + \beta_2 \Gamma_{\!\!2})
= l \} \label{class1}\; ,
\end{eqnarray}
where $\alpha_i, \beta_i \in \mathbb{C} $ and $|\alpha_{i}| +
|\beta_{i}| \neq 0$; $l \in [0, n]$ and $n \in [0, N]$; $r_{\rm
max}$ and $r_{\rm min}$ are maximum and minimum ranks of matrices
for all possible values of $\alpha_i$ and $\beta_i$. From the
definition of (\ref{class1}), there is no common element in
different sets, i.e. $C_{n,\; l} \cap C_{m,\; k} = C_{n,\; l}\;
\delta_{m,n} \delta_{l,k}$. Obviously, every (entangled) state
$(\Gamma_{\!1},\Gamma_{\!\!2})$ must lie in one of the subspaces of
the set $\{C_{n,\;l}\}$ with certain $n$ and $l$, which in principle
can be determined via the transformation of equation (\ref{T-V3}),
since one can always classify a set by certain rules voluntarily.
Here, the criteria $r_{\mathrm{max}}(\alpha_1 \Gamma_{\!1} + \beta_1
\Gamma_{\!\!2}) = n$ and $r_{\mathrm{min}}(\alpha_2 \Gamma_{\!1} +
\beta_2 \Gamma_{\!\!2}) = l$ attribute to the group SL(2, ${\mathbb
C}$) transformation $T$ in (\ref{transfromation-1}). Note that in
case $\left(\begin{array}{ll}
\alpha_1 & \beta_1 \\
\alpha_2 & \beta_2
\end{array}\right)
$ is non-invertible, when $r_{max}$, $r_{min}$, ${\rm
Rank}(\Gamma_1)$, and ${\rm Rank}(\Gamma_2)$ are all equal in
magnitude, its function can be fulfilled by a unit matrix.

Suppose $(\Gamma_{\!1},\ \Gamma_{\!\!2}) \in C_{n,\ l}$,
$(\bar{\Gamma}_{\!1},\ \bar{\Gamma}_{\!\!2})\in C_{\bar{n},\
\bar{l}}$, and
\begin{eqnarray}
{\rm Rank}\left[{\rm O_A}\left(
\begin{array}{l}
\Gamma_{\!1} \\
\Gamma_{\!\!2}
\end{array}
\right)\right] = {\rm Rank}\left[ \left(\begin{array}{ll}
\alpha_1 & \beta_1 \\
\alpha_2 & \beta_2
\end{array}\right)
\left(
\begin{array}{l}
\Gamma_{\!1} \\
\Gamma_{\!\!2}
\end{array}
\right)\right] = {\rm Rank}\left(
\begin{array}{l}
\Gamma'_{\!1} \\
\Gamma'_{\!\!2}
\end{array}
\right) =\left(
\begin{array}{l}
n \\
l
\end{array}
\right) \label{oarank}\; ,
\end{eqnarray}
\begin{eqnarray}
{\rm Rank}\left[{\rm \bar{O}_A}\left(
\begin{array}{l}
\bar{\Gamma}_{\!1} \\
\bar{\Gamma}_{\!\!2}
\end{array}
\right)\right] = {\rm Rank}\left[ \left(\begin{array}{ll}
\bar{\alpha}_1 & \bar{\beta}_1 \\
\bar{\alpha}_2 & \bar{\beta}_2
\end{array}\right)
\left(
\begin{array}{l}
\bar{\Gamma}_{\!1} \\
\bar{\Gamma}_{\!\!2}
\end{array}
\right)\right] = {\rm Rank}\left(
\begin{array}{l}
\bar{\Gamma}'_{\!1} \\
\bar{\Gamma}'_{\!\!2}
\end{array}
\right) =\left(
\begin{array}{l}
\bar{n} \\
\bar{l}
\end{array}
\right) \label{oabrank}\; ,
\end{eqnarray}
where ${\rm O_A}$ and ${\rm \bar{O}_A}$ are invertible operators;
the ``Rank" denotes the rank operation on $\Gamma$ matrices in upper
and lower blocks separately, there will have no invertible matrix
$O_I$ exist, which enables
\begin{eqnarray}
O_I\left(
\begin{array}{l}
\bar{\Gamma}_{\!1} \\
\bar{\Gamma}_{\!\!2}
\end{array}
\right) = \left(
\begin{array}{l}
{\Gamma}_{\!1} \\
{\Gamma}_{\!\!2}
\end{array}
\right) \;\label{oislocc}
\end{eqnarray}
in case $n\neq \bar{n}$ or $l\neq \bar{l}$, i.e. $(\Gamma_{\!1},\
\Gamma_{\!\!2})$  and $(\bar{\Gamma}_{\!1},\ \bar{\Gamma}_{\!\!2})$
are SLOCC inequivalent. If the operator $O_I$ exists, substituting
(\ref{oislocc}) into (\ref{oarank}) we may have
\begin{eqnarray}
{\rm Rank}\left[{\rm {O}_A {O}_I}\left(
\begin{array}{l}
\bar{\Gamma}_{\!1} \\
\bar{\Gamma}_{\!\!2}
\end{array}
\right)\right] = \left(
\begin{array}{l}
{n} \\
{l}
\end{array}
\right) \; ,
\end{eqnarray}
and from (\ref{oabrank}) one knows that $n\leq \bar{n}$ and $l\geq
\bar{l}$. Similarly, since $O_I$ is an invertible operator(matrix),
one may also get $\bar{n} \leq n$ and $\bar{l} \geq l$, and hence,
 $\bar{n} = n$ and $\bar{l} = l$. From the above arguments,
 two states, the matrix pairs
$(\Gamma_{\!1},\ \Gamma_{\!\!2})$ and $(\bar{\Gamma}_{\!1},\
\bar{\Gamma}_{\!\!2})$, connected via invertible operator belong to
the same subset $C_{n,\; l}~$.

Therefore, the entangled classes in set $\{C_{n,\; l}\}$ with
different $n$ and $l$ are SLOCC inequivalent, and the question of
performing a complete classification on entangled states now turns
to how to classify the entangled states in subset $C_{n,\; l}~$.

\subsection{Classification on set $C_{n,l}$ with $n=N$}

From the definition of $C_{N,\,l}$ we know that if
$(\Gamma_{\!1},\Gamma_{\!2})\in C_{N,\,l}$, then there exist an
invertible operator $T$ which enables
\begin{eqnarray}
T \left(
\begin{array}{l}
\Gamma_{\!1} \\
\Gamma_{\!\!2}
\end{array}
\right)  = \left(\begin{array}{ll}
\alpha_1 & \beta_1 \\
\alpha_2 & \beta_2
\end{array}\right)
\left(
\begin{array}{l}
\Gamma_{\!1} \\
\Gamma_{\!\!2}
\end{array}
\right) = \left(
\begin{array}{l}
\Gamma_{\!1}' \\
\Gamma_{\!\!2}'
\end{array}
\right) \; ,
\end{eqnarray}
where $\Gamma_{\!1}'$ has the maximum rank $N$ and $\Gamma_{\!\!2}'$
has minimum rank $l$. According to matrix algebra, in principle one
can find invertible operators $P$, $Q$ and $S$ which further
transform the $(\Gamma_1', \Gamma_2')$ in the following form
\begin{eqnarray}
SP \otimes QS^{-1} \left\lgroup
\begin{array}{c}
\Gamma_{\!1}' \\
\Gamma_{\!\!2}' \\
\end{array}
\right\rgroup \equiv
\left\lgroup
\begin{array}{c}
SP\Gamma_{\!1}'Q S^{-1} \\
SP\Gamma_{\!\!2}'Q S^{-1} \\
\end{array}
\right\rgroup
=
\left\lgroup
\begin{array}{c}
    E \\
    J \\
\end{array}
\right\rgroup \; . \label{jordan}
\end{eqnarray}
Here, $ r(J) = r_{\mathrm{min}}(\alpha_2 \Gamma_{\!1} + \beta_2
\Gamma_{\!\!2}) $ with $J$ a matrix in the Jordan canonical form. A
typical form of $J$ reads
\begin{eqnarray}
J = \left(
  \begin{array}{cccc}
    J_{n_1}(\lambda_{1}) & 0  &  \cdots  & 0   \\
    0 & J_{n_2}(\lambda_{2}) &   & 0  \\
    \vdots  &   & \ddots & \vdots   \\
    0 &  0 & \cdots  & J_{n_k}(\lambda_{k}) \\
  \end{array}
\right)\; , \label{J-L-N-1}
\end{eqnarray}
in which $J_{n_i}(\lambda_i)$ is a $n_i \times n_i$ matrix which has
the following form
\begin{eqnarray}
J_{n_i}(\lambda_i) = \left(
\begin{array}{cccccc}
\lambda_{i} & 1 & 0 & \cdots & 0 & 0 \\
0 & \lambda_{i} & 1 &   & 0 & 0 \\
0 & 0 & \lambda_{i} &   & 0 & 0 \\
\vdots &   &   & \ddots &   &   \\
0 & 0 & 0 &   & \lambda_{i} & 1 \\
0 & 0 & 0 & \cdots & 0 & \lambda_{i} \\
\end{array}
\right) \; . \label{ni-lambdai}
\end{eqnarray}

In all, for every $(\Gamma_{\!1},\Gamma_{\!\!2})\in C_{N,\; l}$,
there exists an ILO transformation, like
\begin{eqnarray}
\left\lgroup
  \begin{array}{c}
    E \\
    J \\
  \end{array}
\right\rgroup = T \otimes P \otimes Q \left\lgroup
\begin{array}{c}
\Gamma_{\!1} \\
\Gamma_{\!\!2} \\
\end{array}
\right\rgroup\; . \label{transf}
\end{eqnarray}
Provided $r(J)=N$, we know that the rank of the matrix $J' \equiv (J
- \lambda_i E)$, with $\lambda_i$ being any eigenvalue of $J$, must
be less than that of $J$'s. This conclusion contradicts with the
proviso of $J$ having the minimum rank, since $J'$ and $J$ are
correlated through an invertible operator, let's say $T'$,
\begin{eqnarray}
\left\lgroup
  \begin{array}{c}
    E \\
    J' \\
  \end{array}
\right\rgroup = T' \left\lgroup
\begin{array}{c}
E \\
J \\
\end{array}
\right\rgroup = \left(\begin{array}{ll}
1 & 0 \\
-\lambda_i & 1
\end{array}\right)
\left\lgroup
\begin{array}{c}
E \\
J \\
\end{array}
\right\rgroup \; .\label{jandjprime}
\end{eqnarray}
From above arguments, one observes that the rank of $J$ is less than
$N$, i.e. $l \leq N-1$. In the special case of $N = 2$, this
observation agrees with the proposition given in Refs.
\cite{inductive-3,local-d}. From Eq.(\ref{transf}) $c_{N,\;l} = (E,
J)$ is equivalent to $C_{N,\;l}$ under the joint invertible
transformations of $T$, $P$, and $Q$, that means the classification
on $C_{N,\ l}$ can be simply performed on $c_{N,\ l}$.

From Eq.(\ref{density-cube-nn}) one can find that for the quantum
state (matrix pair) in $c_{N,l}$
\begin{eqnarray}
\mathrm{Det}(\rho_{\psi_j}) = \prod_i[ \sum_{m=0}^{n_i}
\frac{(1+|\lambda_i|^2)^{m} }{(n_i-m)!}\left. f_{m+1}^{(n_i-m)}(x)
\right|_{x=0}\ ] \neq 0  \label{det-NN}
\end{eqnarray}
with $f_n(x) = \left[ \frac{1 -x}{1 - x - x^2}\right]^n$.  Here,
$j=\psi_1,\psi_2$ and $n_i,\lambda_i$ are defined in
Eq.(\ref{ni-lambdai}). This tells that $r(\rho_{\psi_1}) =
r(\rho_{\psi_2}) = N$. When $l\neq 0$ we readily have
$r(\rho_{\psi_0}) =2$. This means the state in $c_{N, l}$ is true
entangled $2\times N\times N$ state, while $l\neq 0$. Otherwise it
will not be a true entangled $2\times N\times N$ state, which is
beyond our consideration.

\begin{theorem}
$\forall$ $(E,J) \in c_{N,\;l}$, the set $c_{N,\; l}$ is of the
classification of $C_{N,\;l}$ under SLOCC: $\mathrm{(i)}$ if two
states in  $C_{N,\; l}$ are SLOCC equivalent, then they can be
transformed into the same matrix vector $(E, J)$; \\ $\mathrm{(ii)}$
matrix vector $(E, J)$ is unique in $c_{N,\; l}$ up to a trivial
transformation, that is if $(E,J')$ is SLOCC equivalent with
$(E,J)$, then $(E, J') = (E,J + \lambda E)$ with $\lambda$ being an
arbitrary complex number.
\end{theorem}

\noindent
Proof:\\
(i) Suppose there exists the transformation
\begin{eqnarray}
\left\lgroup
  \begin{array}{c}
    \Gamma_{\!1}' \\
    \Gamma_{\!\!2}' \\
  \end{array}
\right\rgroup = T' \otimes P' \otimes Q' \left\lgroup
\begin{array}{c}
\Gamma_{\!1} \\
\Gamma_{\!\!2} \\
\end{array}
\right\rgroup\;,
\end{eqnarray}
according to equation (\ref{transf})
\begin{eqnarray}
\left\lgroup
  \begin{array}{c}
    E \\
    J \\
  \end{array}
\right\rgroup = T \cdot T'^{-1} \otimes P \cdot P'^{-1} \otimes Q
\cdot Q'^{-1} \left\lgroup
  \begin{array}{c}
    \Gamma_{1}' \\
    \Gamma_{\!\!2}' \\
  \end{array}
\right\rgroup\; .
\end{eqnarray}
(ii) Suppose
\begin{eqnarray}
\left\lgroup
  \begin{array}{c}
    E \\
    J' \\
  \end{array}
\right\rgroup = T' \otimes P' \otimes Q'  \left\lgroup
  \begin{array}{c}
    E \\
    J \\
  \end{array}
\right\rgroup\; , \label{N-T}
\end{eqnarray}
as noted beneath the Eq.(\ref{J-L-N-1}) we have $l\leq N-1$, and it
tells that there are no zero elements in the pivot of $T'$. Then,
$T'$ can be decomposed as lower and upper triangular(LU) forms
\cite{linear-algebra}
\begin{eqnarray}
\left\lgroup
  \begin{array}{cc}
    t'_{11} & t'_{12} \\
    t'_{21} & t'_{22} \\
  \end{array}
\right\rgroup = \left\lgroup
  \begin{array}{cc}
    1 & 0 \\
    \lambda & 1 \\
  \end{array}
\right\rgroup \cdot \left\lgroup
  \begin{array}{cc}
    \alpha & \beta \\
    0 & \gamma \\
  \end{array}
\right\rgroup\; ,
\end{eqnarray}
where $\alpha, \beta, \gamma, \lambda \in \mathbb{C}$ and both
matrices on the righthand side are nonsingular. Now Eq.(\ref{N-T})
becomes
\begin{eqnarray}
\left\lgroup
  \begin{array}{c}
    E \\
    J' \\
  \end{array}
\right\rgroup = O_{\!1} O_{2} \left\lgroup
  \begin{array}{c}
    E \\
    J \\
  \end{array}
\right\rgroup\;  \label{J-two}
\end{eqnarray}
with
\begin{eqnarray}
 O_{\!1} = \left\lgroup
  \begin{array}{cc}
    1 & 0 \\
    \lambda & 1 \\
  \end{array}
\right\rgroup,\;\; O_{2} = \left\lgroup
  \begin{array}{cc}
    \alpha & \beta \\
    0 & \gamma \\
  \end{array}
\right\rgroup\otimes  P' \otimes Q' \label{decomposition} \; .
\end{eqnarray}
Here, according to the definition of operator $Q$ in
Eqs.(\ref{T-V3}) and (\ref{jordan}), $Q'$ is applied to matrix
vector from the right hand side. Thus the operator $O_{2}$ acts on
$(E, J)$ in following way
\begin{eqnarray}
O_{2} \left\lgroup
  \begin{array}{c}
    E \\
    J \\
  \end{array}
\right\rgroup & = & P' \left\lgroup
  \begin{array}{cc}
    \alpha & \beta \\
    0 & \gamma \\
  \end{array}
\right\rgroup  \left\lgroup
  \begin{array}{c}
    E \\
    J \\
  \end{array}
\right\rgroup Q' \; , \label{V1-rescale}
\end{eqnarray}
where $J =J(\lambda_i) =  \oplus_i J_{n_i}(\lambda_{i})$.  The
Eq.(\ref{J-two}) now gives two independent equations
\begin{eqnarray}
E & = & P'(\alpha E + \beta J) Q' \; ,\label{sovlve-pq}\\
J' & = & \lambda E + \gamma P'J Q' \; .
\end{eqnarray}
The first one can be reformulated into a different form
\begin{eqnarray}
P'  J(\alpha+\beta \lambda_i)Q' & = & E \; ,  \label{J-inverse}
\end{eqnarray}
and from eqs.(\ref{sovlve-pq}) and (\ref{J-inverse}) we can further
get
\begin{eqnarray}
\frac{\beta}{\alpha} P' J Q' & = & \frac{1}{\alpha} E - P'Q'
\nonumber \\ & = & \frac{1}{\alpha} E - Q'^{-1}J^{-1}(\alpha +
\beta \lambda_i) Q' \nonumber \\
& = & Q'^{-1}M^{-1}( \frac{1}{\alpha} E - J(\frac{1}{\alpha +
\beta\lambda_{i}}))MQ' \nonumber \\ & = &
Q'^{-1}M^{-1}J(\frac{\beta\lambda_{i}}{\alpha(\alpha +
\beta\lambda_{i})})MQ'\; .
\end{eqnarray}
Here, $M$ is an invertible matrix, and the theorem (6.2.25) in
\cite{r.a.Horn2} is employed. Thus,
\begin{eqnarray}
\gamma P'J Q' & = & Q'^{-1} M^{-1} J
(\frac{\gamma\lambda_{i}}{\alpha + \beta\lambda_{i}}) MQ' \nonumber
\\ & = & S J (\frac{\gamma\lambda_{i}}{\alpha + \beta\lambda_{i}})
S^{-1} \label{add1}\;
\end{eqnarray}
with $S = Q'^{-1}M^{-1} $. Therefore, from
(\ref{V1-rescale})-(\ref{add1}) we get
\begin{eqnarray}
O_{2} \left\lgroup
  \begin{array}{c}
    E \\
    J \\
  \end{array}
\right\rgroup
 =  \left\lgroup
\begin{array}{c} E \\
S \oplus_{i}J_{n_i}(\frac{\gamma \lambda_{i}}{\alpha + \beta \lambda_{i}})S^{-1} \\
\end{array}
\right\rgroup \; , \label{reV1-rescale}
\end{eqnarray}
and hence
\begin{eqnarray}
\left\lgroup
  \begin{array}{c}
    E \\
    J' \\
  \end{array}
\right\rgroup = \left\lgroup
\begin{array}{cc}
1 & 0 \\
\lambda & 1 \\
\end{array}
\right\rgroup \left\lgroup
\begin{array}{c}
E \\
J \\
\end{array}
\right\rgroup\; .
\end{eqnarray}
Q.E.D.

\subsection{Classification on set $C_{n,l}$ with
$n=N-1$}\label{section-theorem2}
Having classified the $C_{N,\,l}$,
now we proceed to the $C_{N-1,\,l}$ case. For every
$(\Gamma_{\!1},\Gamma_{\!\!2})\in C_{N-1,\; l}$, we can find an ILO
$T$ which implements the following transformation
\begin{eqnarray}
\left\lgroup
\begin{array}{ll}
t_{11} & t_{12} \\
t_{21} & t_{22}
\end{array}
\right\rgroup
\left\lgroup
\begin{array}{c}
\Gamma_{\!1} \\
\Gamma_{\!\!2} \\
\end{array}
\right\rgroup =
\left\lgroup
\begin{array}{c}
\Gamma_{\!1}' \\
\Gamma_{\!\!2}' \\
\end{array}
\right\rgroup \; ,
\end{eqnarray}
where $r(\Gamma_{\!1}') = N-1$ and $r(\Gamma_{\!\!2}') = l$. Then in
principle one can find ILOs $P_1$ and $Q_1$, which transform the
matrix vector $(\Gamma_{\!1}', \Gamma_{\!\!2}')$ into the form
$(\Lambda , \Gamma_{\!\!2}'')$ where $\Lambda$ is a $N\times N$
diagonal matrix with N-1 nonzero elements of 1 and one zero, and $
\Gamma_{\!\!2}''$ is another $N\times N$ matrix in a specific form.
To be more explicit, taking $N=6$ as an example (but the procedure
is $N$ independent) the above mentioned procedure tells
\begin{eqnarray}
\Lambda = P \Gamma_{\!1}' Q & = & \left(
\begin{array}{llllll}
1 & 0 & 0 & 0 & 0 & 0 \\
0 & 1 & 0 & 0 & 0 & 0 \\
0 & 0 & 1 & 0 & 0 & 0 \\
0 & 0 & 0 & 1 & 0 & 0 \\
0 & 0 & 0 & 0 & 1 & 0 \\
0 & 0 & 0 & 0 & 0 & 0
\end{array}
\right)\; , \label{N-1-20} \\
\Gamma_{\!2}'' = P \Gamma_{\!2}' Q & = & \left(
  \begin{array}{ccc|ccc}
    \times & \times & \times & 0 & \times & 0 \\
    \times & \times & \times & 0 & \times & 0 \\
    \times & \times & \times & 0 & \times & 0 \\ \hline
    \times & \times & \times & 0 & 0 & 0 \\
    0 & 0 & 0 & 0 & 0 & 1 \\
    0 & 0 & 0 & 1 & 0 & 0 \\
  \end{array}
\right) =
\left(
\begin{array}{cc}
A & c \\
r & B_{3} \\
\end{array}
\right)\; ,  \label{N-1-2}
\end{eqnarray}
where $A$, $B_3$, $c$ and $r$ are submatrices of $\Gamma_{\!2}''$
partitioned by vertical and horizontal lines; $\times$ means no
constraint on the corresponding element. Note that for the case of
N-by-N matrix,  $\Gamma_{\!2}''$ can also be partitioned into a
similar form as (\ref{N-1-2}), with the lower right block to be
still the $B_3$ (see Appendix \ref{construct-B-matrix} for details).

In general, there are four different choices for $c$ and $r$, i.e.,
1) $ c = 0,\ r = 0$; 2) $c\neq 0,\ r = 0$; 3) $c= 0,\ r\neq 0$; 4)
$c\neq 0,\ r\neq 0$. In these cases $\Gamma_{\!2}''$ can be further
simplified through a series of elementary operations, e.g. denoted
$P_{k}$ and $Q_{k}$, which enables
\begin{eqnarray}
\left\lgroup
\begin{array}{l}
\Lambda \\
\Gamma_{\!\!2}''^{cr}
\end{array}
\right\rgroup =
\left\lgroup
\begin{array}{l}
P_k\Lambda Q_k \\
P_k\Gamma_{\!\!2}''Q_k
\end{array}
\right\rgroup\; . \label{transf-t2}
\end{eqnarray}
Here, the superscripts $c$ and $r$ stand for different choices
mentioned in above, and the $\Gamma_{\!2}''^{cr}$ read
\begin{eqnarray}
\Gamma_{\!2}''^{00} & = & \left(
  \begin{array}{cccccc}
    \times & \times & \times & 0 & 0 & 0 \\
    \times & \times & \times & 0 & 0 & 0 \\
    \times & \times & \times & 0 & 0 & 0 \\
    0 & 0 & 0 & 0 & 0 & 0 \\
    0 & 0 & 0 & 0 & 0 & 1 \\
    0 & 0 & 0 & 1 & 0 & 0 \\
  \end{array}
\right) ,\ \ \Gamma_{\!2}''^{10} =  \left(
  \begin{array}{ccc|ccc}
    \times & \times & \times & 0 & 0 & 0 \\
    \times & \times & \times & 0 & 0 & 0 \\
    0 & 0 & 0 & 0 & 1 & 0 \\ \hline
    0 & 0 & 0 & 0 & 0 & 0 \\
    0 & 0 & 0 & 0 & 0 & 1 \\
    0 & 0 & 0 & 1 & 0 & 0 \\
  \end{array}
\right) ,\ \ \nonumber \\[0.3cm]
\Gamma_{\!2}''^{01} & = & \left(
  \begin{array}{ccc|ccc}
    \times & \times & 0 & 0 & 0 & 0 \\
    \times & \times & 0 & 0 & 0 & 0 \\
    \times & \times & 0 & 0 & 0 & 0 \\ \hline
    0 & 0 & 1 & 0 & 0 & 0 \\
    0 & 0 & 0 & 0 & 0 & 1 \\
    0 & 0 & 0 & 1 & 0 & 0 \\
  \end{array}
\right) ,\ \ \Gamma_{\!2}''^{11}  =  \left(
  \begin{array}{ccc|ccc}
    \times & 0 & \times & 0 & 0 & 0 \\
    \times & 0 & 0 & 0 & 0 & 0 \\
    0 & 0 & 0 & 0 & 1 & 0 \\ \hline
    0 & 1 & 0 & 0 & 0 & 0 \\
    0 & 0 & 0 & 0 & 0 & 1 \\
    0 & 0 & 0 & 1 & 0 & 0 \\
  \end{array}
\right)\; . \label{redf}
\end{eqnarray}
In the case of $\Gamma_{\!2}''^{00}$, it has
already been in the form of $\left(\begin{array}{cc}A & 0 \\ 0 & B_3 \\
\end{array} \right)$. In the other three cases we can repartition
the submatrices, like
\begin{eqnarray}
\Gamma_{\!2}''^{10} =  \left(
  \begin{array}{cc|cccc}
    \times & \times & \times & 0 & 0 & 0 \\
    \times & \times & \times & 0 & 0 & 0 \\ \hline
    0 & 0 & 0 & 0 & 1 & 0 \\
    0 & 0 & 0 & 0 & 0 & 0 \\
    0 & 0 & 0 & 0 & 0 & 1 \\
    0 & 0 & 0 & 1 & 0 & 0 \\
  \end{array}
\right) = \left(
\begin{array}{cc}
A & c \\
0 & B_{4} \\
\end{array}
\right) \; , \label{gamma-10}
\end{eqnarray}
\begin{eqnarray} \Gamma_{\!2}''^{01}   =
\left(
  \begin{array}{cc|cccc}
    \times & \times & 0 & 0 & 0 & 0 \\
    \times & \times & 0 & 0 & 0 & 0 \\ \hline
    \times & \times & 0 & 0 & 0 & 0 \\
    0 & 0 & 1 & 0 & 0 & 0 \\
    0 & 0 & 0 & 0 & 0 & 1 \\
    0 & 0 & 0 & 1 & 0 & 0 \\
  \end{array}
\right) = \left(
\begin{array}{cc}
A & 0 \\
r & B_{4} \\
\end{array}
\right) \; , \label{gamma-01}
\end{eqnarray}
\begin{eqnarray}
\Gamma_{\!2}''^{11}  = \left(
  \begin{array}{c|ccccc}
    \times & 0 & \times & 0 & 0 & 0 \\ \hline
    \times & 0 & 0 & 0 & 0 & 0 \\
    0 & 0 & 0 & 0 & 1 & 0 \\
    0 & 1 & 0 & 0 & 0 & 0 \\
    0 & 0 & 0 & 0 & 0 & 1 \\
    0 & 0 & 0 & 1 & 0 & 0 \\
  \end{array}
\right) = \left(
\begin{array}{cc}
A & c \\
r & B_{5} \\
\end{array}
\right) \; , \label{gamma-11}
\end{eqnarray}
where $n$ in $B_{n}$ means the dimension of the matrix. This
procedure results in the enlargement of the blocks $B_n$, the shrink
of blocks $A$, and the emergence of new types of off diagonal blocks
$c$ and $r$.

By the same procedure, one can further simplify, enlarge $B_n$ and
shrink $A$, the $\Gamma_{\!2}''^{cr}$ of the forms
(\ref{gamma-10}-\ref{gamma-11}). Finally  the $\Gamma_{\!\!2}''$ in
(\ref{N-1-2}) may arrive at the form of
\begin{eqnarray}
\Gamma_{\!2}'' \sim \left(
  \begin{array}{cc}
    A & 0 \\
    0 & B_n \\
  \end{array}
\right) =
\left(
  \begin{array}{cc}
    D J D^{-1} & 0 \\
    0 & B_n \\
  \end{array}
\right) \label{D-jordan}
\end{eqnarray}
with different kinds of $B_n$s, correspondingly. Here, $J$ is the
Jordan canonical form of $A$, and $D$ is an invertible matrix. Every
$\Gamma_{\!\!2}''$ matrix has its own specific form of $B_n$ block.
Note that $B_n$ may be recursively enlarged to be the whole matrix
of $\Gamma_{\!2}''$, i.e., $n=N$. In all, for every
$(\Gamma_{\!1},\Gamma_{\!\!2})\in C_{N-1,\, l}$, there exists an ILO
transformation, like
\begin{eqnarray}
\left\lgroup
  \begin{array}{c}
    \Lambda \\
    \Gamma \\
  \end{array}
\right\rgroup = T \otimes P \otimes Q
\left\lgroup
\begin{array}{c}
\Gamma_{\!1} \\
\Gamma_{\!\!2} \\
\end{array}
\right\rgroup\; . \label{transf-n-1}
\end{eqnarray}
Here, the $ \Gamma = \left(\begin{array}{cc} J & 0 \\0 & B_n \\
\end{array} \right)$, $P=\prod_{i} P_{i}$ and $Q = \prod_{i}Q_{i}$,
where $P_i, Q_i$ stand for those operators in Eqs. (\ref{N-1-2}),
(\ref{transf-t2}), and (\ref{D-jordan}). From Eq.(\ref{transf-n-1})
the subset $c_{N-1,\;l}$, defined as
\begin{eqnarray}
c_{N-1,\;l}=\{ (\Lambda, \Gamma)|\; r(\Gamma) = l; \Gamma =
\left(\begin{array}{cc} J & 0 \\0 & B_n
\\ \end{array} \right); \; (\Lambda,
\Gamma)\in C_{N-1,\; l} \} \; ,
\end{eqnarray}
is equivalent to $C_{N-1,\;l}$ under the joint invertible
transformations of $T$, $P$, and $Q$, which means that the
classification on $C_{N-1,\ l}$ can be simply performed on $c_{N-1,\
l}$.

Similar as ( \ref{det-NN}) we find
\begin{eqnarray} \mathrm{Det}
(\rho_{j}) = \prod_i[ \sum_{m=0}^{n_i} \frac{(1+|\lambda_i|^2)^{m}
}{(n_i-m)!}\left. f_{m+1}^{(n_i-m)}(x) \right|_{x=0}\ ] \cdot
2^{\,r(B_n)-1} \neq 0 \; ,
\end{eqnarray}
where $j=\psi_1,\psi_2$, and $n_i,\lambda_i$ is defined as the
Jordan blocks in Eq.(\ref{D-jordan}), and conclude that all the
states in $c_{N-1,l}$ are truly entangled in the form of $2\times
N\times N$.

\begin{theorem}

$\forall$ $(\Lambda, \Gamma) \in c_{N-1,\;l}$, the set $c_{N-1,\;
l}$ is of the classification of $C_{N-1,\;l}$ under SLOCC: \\
$\mathrm{(i)}$ suppose two states in $C_{N-1,\;l}$ are SLOCC
equivalent, they can then be transformed
into the same matrix vector $(\Lambda, \Gamma)$; \\
$\mathrm{(ii)}$ the matrix vector in $c_{N-1,\,l}$ is unique up to a
nonzero classification irrelevant parameter, i.e., provided
$(\Lambda,\Gamma')$ is SLOCC equivalent with $(\Lambda,\Gamma)$,
then $(\Lambda, \Gamma') = (\Lambda, \kappa\Gamma)$ in the sense of
$J'$ equalling to $J$ as in theorem 1 while $B_n'$ being exactly the
same as $B_n$.
\end{theorem}

\noindent
Proof: \\
(i) According to the property of transitivity in SLOCC
transformation, this proposition should be true.

\noindent
(ii) Suppose
\begin{eqnarray}
\left(
  \begin{array}{c}
    \Lambda \\
    \Gamma' \\
  \end{array}
\right) = T' \otimes P' \otimes Q' \left(
  \begin{array}{c}
    \Lambda \\
    \Gamma \\
  \end{array}
\right)\label{trans54}\; ,
\end{eqnarray}
we first demonstrate that the three ILOs transformation $T',P',Q'$
can always be fulfilled through two ILOs transformations $P'',Q''$,
that is
\begin{eqnarray}
T' \otimes P' \otimes Q' \left(
  \begin{array}{c}
    \Lambda \\
    \Gamma \\
  \end{array}
\right) & = & P'' \otimes Q'' \left(
  \begin{array}{c}
    \Lambda \\
   \kappa \Gamma \\
  \end{array}
\right) \nonumber \\ & = & \left(
  \begin{array}{c}
   P'' \Lambda Q''\\
   P'' \kappa\Gamma Q'' \\
  \end{array}
\right) \; . \label{three-two-tran}
\end{eqnarray}

According to the definition of $c_{N-1,\, l}$ we can write
$(\Lambda, \Gamma)$ in the form of direct sums
\begin{eqnarray}
\left(
\begin{array}{c}
\Lambda \\
\Gamma \\
\end{array}
\right) = \left(
\begin{array}{c}
\left(
\begin{array}{cc}
E & 0 \\
0 & \Lambda' \\
\end{array}
\right)
\\
\left(
\begin{array}{cc}
J & 0 \\
0 & B_n \\
\end{array}
\right)
\\
\end{array}
\right)\; , \label{dirct-sums}
\end{eqnarray}
where
\begin{eqnarray}
\Lambda' =
\left(
\begin{array}{lllll}
1 & 0 & \cdots & 0 & 0 \\
0 & 1 &   & 0 & 0 \\
\vdots &  & \ddots &    & \vdots  \\
0 & 0 & \cdots & 1 & 0 \\
0 & 0 & \cdots & 0 & 0
\end{array}
\right)
\end{eqnarray}
is a square matrix, and its dimension equals to that of $B_n$. The
transformation  $T'$
\begin{eqnarray}
\left\lgroup
  \begin{array}{cc}
    1 & 0 \\
    \lambda & 1 \\
  \end{array}
\right\rgroup
\left\lgroup
  \begin{array}{cc}
    \alpha & \beta \\
    0 & \gamma \\
  \end{array}
\right\rgroup
\left(
\begin{array}{c}
 \Lambda \\
 \Gamma  \\
\end{array}
\right)\; \label{self-t}
\end{eqnarray}
can be decomposed into the following form
\begin{eqnarray}
\left\lgroup
  \begin{array}{cc}
    1 & 0 \\
    \lambda & 1 \\
  \end{array}
\right\rgroup
\left\lgroup
  \begin{array}{cc}
    \alpha & \beta \\
    0 & \gamma \\
  \end{array}
\right\rgroup
\left(
\begin{array}{c}
 E  \\
 J \\
\end{array}
\right)\; , \label{J-part} \\
\left\lgroup
  \begin{array}{cc}
    1 & 0 \\
    \lambda & 1 \\
  \end{array}
\right\rgroup
\left\lgroup
  \begin{array}{cc}
    \alpha & \beta \\
    0 & \gamma \\
  \end{array}
\right\rgroup
\left(
\begin{array}{c}
 \Lambda'  \\
 B_n  \\
\end{array}
\right)\; , \label{B-part}
\end{eqnarray}
according to the nature of direct sum.


For Eq.(\ref{J-part}), from the proof of {\it theorem 1} one can
find the following $P_{J}, Q_{J}$
\begin{eqnarray}
\left\lgroup
  \begin{array}{cc}
    1 & 0 \\
    \lambda & 1 \\
  \end{array}
\right\rgroup
\left\lgroup
  \begin{array}{cc}
    \alpha & \beta \\
    0 & \gamma \\
  \end{array}
\right\rgroup
\left(
\begin{array}{c}
P_{J} E Q_{J} \\
P_{J} J Q_{J} \\
\end{array}
\right) =
\left(
\begin{array}{c}
 E \\
 J \\
\end{array}
\right)\; ,
\end{eqnarray}
should exist. Here $J + \lambda E$ is taken to be equivalent to the
$J$ while $r(J + \lambda E) = r(J) = l$.

For Eq.(\ref{B-part}), we take into account the two decomposed
operations of $T'$ separately. We have
\begin{eqnarray}
\left(
  \begin{array}{cc}
    \alpha & \beta \\
    0 & \gamma \\
  \end{array}
\right)\left(
\begin{array}{c}
 \Lambda'  \\
 B_n  \\
\end{array}
\right) = \left(
\begin{array}{c}
 \alpha(\Lambda' + \sigma B_n)  \\
 \gamma B_n  \\
\end{array}
\right) = \left(
  \begin{array}{cc}
    \alpha & 0 \\
    0 & \gamma \\
  \end{array}
\right) \left(
\begin{array}{c}
 \Lambda' + \sigma B_n  \\
 B_n  \\
\end{array}
\right) \;
\end{eqnarray}
with $\sigma =\frac{\beta}{\alpha}~$. As shown in Appendix
\ref{superposition-lambda-gamma}, there exist ILOs $P_{Bn}$ and
$Q_{Bn}$ which satisfy
\begin{eqnarray}
P_{Bn}\left(
\begin{array}{c}
 \Lambda' + \lambda' B_n  \\
 B_n  \\
\end{array}
\right)Q_{Bn} = \left(
\begin{array}{c}
 \Lambda' \\
 B_n  \\
\end{array}
\right)\; . \label{app-prove}
\end{eqnarray}
And, further more we have
\begin{eqnarray}
\left\lgroup
  \begin{array}{cc}
    1 & 0 \\
    \lambda & 1 \\
  \end{array}
\right\rgroup \left(
  \begin{array}{cc}
    \alpha & 0 \\
    0 & \gamma \\
  \end{array}
\right) \left(
\begin{array}{c}
 \Lambda' \\
 B_n \\
\end{array}
\right)= \left(
  \begin{array}{cc}
    \alpha & 0 \\
    0 & \gamma \\
  \end{array}
\right) \left(
\begin{array}{c}
 \Lambda' \\
 B_n + \frac{\alpha\lambda}{\gamma} \Lambda' \\
\end{array}
\right) \; .
\end{eqnarray}
There also exist ILOs $P'_{Bn}$ and $Q'_{Bn}$ which satisfy (see
Appendix \ref{superposition-lambda-gamma})
\begin{eqnarray}
P'_{Bn}\left(
\begin{array}{c}
 \Lambda' \\
 B_n + \lambda \Lambda' \\
\end{array}
\right)Q'_{Bn} = \left(
\begin{array}{c}
 \Lambda'   \\
 B_n  \\
\end{array}
\right)\; . \label{Appendix-B}
\end{eqnarray}
Thus Eq.(\ref{B-part}) becomes
\begin{eqnarray}
\left\lgroup
  \begin{array}{cc}
    1 & 0 \\
    \lambda & 1 \\
  \end{array}
\right\rgroup
\left\lgroup
  \begin{array}{cc}
    \alpha & \beta \\
    0 & \gamma \\
  \end{array}
\right\rgroup
\left(
\begin{array}{c}
 P_{B}\Lambda'Q_{B}  \\
 P_{B} B_n Q_{B}  \\
\end{array}
\right) =
 \left(
  \begin{array}{cc}
    \alpha & 0 \\
    0 & \gamma \\
  \end{array}
\right) \left(
\begin{array}{c}
 \Lambda'  \\
 B_n \\
\end{array}
\right) \; ,
\end{eqnarray}
where $P_{B} = P'_{Bn}P_{Bn}$ and $Q_{B} = Q_{Bn}Q'_{Bn}$. By taking
$P_{0}=P_{J}\oplus \frac{1}{\alpha}P_{B}$ and $Q_{0}= Q_{J} \oplus
Q_{B}$, we have
\begin{eqnarray}
&&  \left\lgroup
  \begin{array}{cc}
    1 & 0 \\
    \lambda & 1 \\
  \end{array}
\right\rgroup
\left\lgroup
  \begin{array}{cc}
    \alpha & \beta \\
    0 & \gamma \\
  \end{array}
\right\rgroup
\left(
\begin{array}{c}
P_{0} \Lambda Q_{0}\\
P_{0} \Gamma Q_{0} \\
\end{array}
\right) = \left(
\begin{array}{c}
\left(
\begin{array}{cc}
E & 0 \\
0 & \Lambda' \\
\end{array}
\right)
\\
\left(
\begin{array}{cc}
J & 0 \\
0 & \frac{\gamma}{\alpha}B_n \\
\end{array}
\right)
\\
\end{array}
\right)\nonumber\\
= &&\left(
\begin{array}{c}
\left(
\begin{array}{cc}
E & 0 \\
0 & \Lambda' \\
\end{array}
\right)
\\
\kappa\left(
\begin{array}{cc}
\frac{1}{\kappa}J & 0 \\
0 & B_n \\
\end{array}
\right)
\\
\end{array}
\right) = \left(
  \begin{array}{c}
    \Lambda \\
    \kappa \Gamma \\
  \end{array}
\right) \; .
\end{eqnarray}
Then, Eq.(\ref{trans54}) can be expressed as
\begin{eqnarray}
\left(
  \begin{array}{c}
    \Lambda \\
    \Gamma' \\
  \end{array}
\right) & = & P'\left\lgroup
  \begin{array}{cc}
    1 & 0 \\
    \lambda & 1 \\
  \end{array}
\right\rgroup \left\lgroup
  \begin{array}{cc}
    \alpha & \beta \\
    0 & \gamma \\
  \end{array}
\right\rgroup \left(
  \begin{array}{c}
   \Lambda \\
    \Gamma \\
  \end{array}
\right) Q'\nonumber \\ & = & P'P_{0}^{-1} \left(
  \begin{array}{c}
   \Lambda \\
   \kappa \Gamma \\
  \end{array}
\right) Q_{0}^{-1}Q' \nonumber \\ & = &  \left(
  \begin{array}{c}
  P'' \Lambda Q'' \\
  P'' \kappa \Gamma Q'' \\
  \end{array}
\right) \; , \label{prove-3-2}
\end{eqnarray}
which is just Eq.(\ref{three-two-tran}).

Eq.(\ref{prove-3-2}) corresponds to two equations
\begin{eqnarray}
\left\{ \begin{array}{c}  P'' \Lambda Q'' =  \Lambda \\ \\ P''
\kappa\Gamma Q'' = \Gamma'
\end{array} \right. \; . \label{eqns-constraint}
\end{eqnarray}
The equation $\Lambda = P'' \Lambda Q''$ requires $P''$ and $Q''$
taking the following form
\begin{eqnarray}
P'' = \left(
  \begin{array}{cc}
    S & Y \\
    0 & p \\
  \end{array}
\right);\; Q'' = \left(
  \begin{array}{cc}
    S^{-1} & 0 \\
    X & q \\
  \end{array}
\right)\; , \label{constrain-lam}
\end{eqnarray}
respectively. Here, $p$ and $q$ are arbitrary complex numbers. Note
that in order to guarantee $P''$ and $Q''$ to be non-singular
matrices, $p$ and $q$ can not be zero.

The canonical form of $\Gamma'$ in $c_{N-1,l}$ gives further
constraints on the patterns of $P''$ and $Q''$. Of the $\Gamma'$ and
$\Gamma$, in which the $B_3$ takes the form of (\ref{N-1-2}), they
can be generally expressed as
\begin{eqnarray}
\left(
  \begin{array}{cccccc}
    \times & \times & \times & 0 & v_1 & 0 \\
    \times & \times & \times & 0 & v_2 & 0 \\
    \times & \times & \times & 0 & v_3 & 0 \\
    \times & \times & \times & 0 & 0 & 0 \\
    0 & 0 & 0 & 0 & 0 & 1 \\
    0 & 0 & 0 & 1 & 0 & 0 \\
  \end{array}
\right) \label{constrain-gam}\; .
\end{eqnarray}
In fact any elements in set $c_{N-1,l}$ takes the same patten
(\ref{constrain-gam}). From (\ref{constrain-lam}), $P''$ and $Q''$
take the forms of
\begin{eqnarray}
P''=\left(
  \begin{array}{cccccc}
    a_{11} & a_{12} & a_{13} & a_{14} & a_{15} & y_1 \\
    a_{21} & a_{22} & a_{23} & a_{24} & a_{25} & y_2 \\
    a_{31} & a_{32} & a_{33} & a_{34} & a_{35} & y_3 \\
    a_{41} & a_{42} & a_{43} & a_{44} & a_{45} & y_4 \\
    a_{51} & a_{52} & a_{53} & a_{54} & a_{55} & y_5 \\
    0 & 0 & 0 & 0 & 0 & p \\
  \end{array}
\right),\; Q''=\left(
  \begin{array}{cccccc}
    b_{11} & b_{12} & b_{13} & b_{14} & b_{15} & 0 \\
    b_{21} & b_{22} & b_{23} & b_{24} & b_{25} & 0 \\
    b_{31} & b_{32} & b_{33} & b_{34} & b_{35} & 0 \\
    b_{41} & b_{42} & b_{43} & b_{44} & b_{45} & 0 \\
    b_{51} & b_{52} & b_{53} & b_{54} & b_{55} & 0 \\
    x_1 & x_2 & x_3 & x_4 & x_5 & q\\
  \end{array}
\right)\; , \label{more}
\end{eqnarray}
where $S = \{a_{ij}\}_{5\times 5} = \{b_{ij}\}_{5\times 5}^{-1}$.
From (\ref{constrain-gam}) and constraint $\Gamma' = P'' \kappa
\Gamma Q''$, $P''$ and $Q''$ read
\begin{eqnarray}
P''=\left(
  \begin{array}{cccccc}
    a_{11} & a_{12} & a_{13} & a_{14} & 0 & y_1 \\
    a_{21} & a_{22} & a_{23} & a_{24} & 0 & y_2 \\
    a_{31} & a_{32} & a_{33} & a_{34} & 0 & y_3 \\
    0 & 0 & 0 & p\kappa & 0 & y_4 \\
    a_{51} & a_{52} & a_{53} & a_{54} & 1/q\kappa & y_5 \\
    0 & 0 & 0 & 0 & 0 & p \\
  \end{array}
\right),\; Q''=\left(
  \begin{array}{cccccc}
    b_{11} & b_{12} & b_{13} & b_{14} & 0 & 0 \\
    b_{21} & b_{22} & b_{23} & b_{24} & 0 & 0 \\
    b_{31} & b_{32} & b_{33} & b_{34} & 0 & 0 \\
    0 & 0 & 0 & 1/p\kappa & 0 & 0 \\
    b_{51} & b_{52} & b_{53} & b_{54} & q\kappa & 0 \\
    x_1 & x_2 & x_3 & x_4 & x_5 & q \\
  \end{array}
\right)\; .
\end{eqnarray}
We notice that if $P''$ and $Q''$ are invertible, then the
upper-left sub-matrices $\{a_{ij}\}_{3\times 3}$ and
$\{b_{ij}\}_{3\times 3}$ should also be invertible, due to the fact
that any invertible matrix in the form
$\left(
\begin{array}{cc}
X & 0 \\
W & Y \\
\end{array}
\right)$ has an inverse $\left(
\begin{array}{cc}
X^{-1} & 0 \\
Z & Y^{-1} \\
\end{array}
\right)$ with arbitrary matrices $X$ and $Y$ being also invertible.
From (\ref{eqns-constraint}) one may infer that in
$P''\,\kappa\Gamma Q''$ only the block $\{a_{ij}\}_{3\times 3}$ acts
on vector $v=\{v_i, i=1,2,3\}$ in $\Gamma$ of (\ref{constrain-gam}).
Since no invertible operator can transform a nonzero vector into a
null one, therefore $v=0$ and $v\neq 0$ determine two ILO
inequivalent cases for $\Gamma$. Thus we see that if $\Gamma'$ and
$\Gamma$ in $c_{N-1,l}$ satisfy Eq.(\ref{eqns-constraint}), the
identity of their $B_3$ sub-matrices leads to the identity of their
$B_4$ sub-matrices. Or in other words, for two matrices $\Gamma$ and
$\Gamma'$, if their sub-matrices $B_3$ are the same, while their
$B_4$ sub-matrices are different, then they should be ILO
inequivalent, like (\ref{gamma-10}) and (\ref{gamma-01}).

The above analysis for $B_3$ is applicable to other sub-matrices
$B_n$ with $n > 3$, e.g., the forms of $B_4$ and $B_5$ in
$\Gamma_{\!\!2}^{''}$ in Eqs.(\ref{gamma-10})-(\ref{gamma-11}).
Generally every nonzero element in $B_{n}$ will transform one column
or one row of $P''$ and $Q''$ into a unit vector. In the end, under
the constraint (\ref{eqns-constraint}), if two matrices $\Gamma$ and
$\Gamma'$ have the same $B_i$, then they must possess the same
$B_{i+1}$. According to theorem 1 we then have
\begin{eqnarray}
P'' \kappa \left(\begin{array}{cc} J(\lambda_i) &
0 \\ 0 & B_n \\
\end{array} \right) Q'' = \left(\begin{array}{cc} S J(\kappa\lambda_i) S^{-1} & 0
\\ 0 & B_n \\ \end{array} \right)\; ,
\end{eqnarray}
that is
\begin{eqnarray}
\left\lgroup
  \begin{array}{c}
    \Lambda\\
    \Gamma' \\
  \end{array}
\right\rgroup = \left\lgroup
  \begin{array}{c}
    \Lambda \\
    \Gamma \\
  \end{array}
\right\rgroup\;.
\end{eqnarray}
(Notice that there exists a special case in which $B$ matrix takes
up the whole $\Gamma$ and then the (\ref{J-part}) does not exist.
The elements in the pivot of $T'$ now can have zeros. In this
situation, the only modification of the above proofing process is by
adding two more ILOs $P_{t}$ and $Q_{t}$  which can reverse the flip
of the $(\Lambda', B)$ induced by the zero elements in the pivot of
$T'$, see Appendix \ref{flip-L-B} for details)\\
Q.E.D.

\subsection{Classification on set $C_{n,l}$ with $n=N-i$}

The same procedure can be directly applied to the $C_{N-2,\,l}$
case, and so on. Taken here again the $N=6$ case as an example,
following the construction procedure in Eq.(\ref{N-1-2}) it is easy
to obtain the canonical form of $(\Lambda,\Gamma)$ in $c_{6-2,\,l}$
\begin{eqnarray}
\Lambda & = &
\left(
\begin{array}{ll|ll|ll}
1 & 0 & 0 & 0 & 0 & 0 \\
0 & 1 & 0 & 0 & 0 & 0 \\ \hline
0 & 0 & 1 & 0 & 0 & 0 \\
0 & 0 & 0 & 1 & 0 & 0 \\ \hline
0 & 0 & 0 & 0 & 0 & 0 \\
0 & 0 & 0 & 0 & 0 & 0
\end{array}
\right)\; , \label{N-2-4-lambda} \\
\Gamma & = & \left(
  \begin{array}{cc|cc|cc}
    0 & 0 & 0 & 0 & 0 & 0 \\
    0 & 0 & 0 & 0 & 0 & 0 \\ \hline
    0 & 0 & 0 & 0 & 1 & 0 \\
    0 & 0 & 0 & 0 & 0 & 1 \\ \hline
    1 & 0 & 0 & 0 & 0 & 0 \\
    0 & 1 & 0 & 0 & 0 & 0 \\
  \end{array}
\right) \label{N-2-4-gamma}\; .
\end{eqnarray}
In this case, it is obvious that $l$ can not be smaller than 4.
Rescale the matrices (\ref{N-2-4-lambda}) and (\ref{N-2-4-gamma})
according to the partition lines we have
\begin{eqnarray}
\Lambda & = & \left(
\begin{array}{lll}
E & 0 & 0 \\
0 & E & 0 \\
0 & 0 & 0
\end{array}
\right)\; , \\
\Gamma & = & \left(
  \begin{array}{ccc}
    0 &  0 & 0 \\
    0 &  0 & E \\
    E &  0 & 0 \\
  \end{array}
\right) \; .
\end{eqnarray}
Here $\Gamma$ is just like $B_3$ in Eq.(\ref{N-1-2}). Thus the
classification of set $c_{N-i,l}$ with $i >1$ can be reduced to the
classification of $c_{N-1,l}$ in principle. From (\ref{N-1-2}) and
(\ref{N-2-4-gamma}) we notice that a proposition(inequality) of
$n,l$ in $c_{n,\,l}$ should exist, i.e.
\begin{eqnarray}
 2(N-n)\leq l \leq n \label{nl-prop}\; ,
\end{eqnarray}
when $n < N$. Eq.(\ref{nl-prop}) is a constraint on the rank of
matrix-pair which represents the true entangled state of $2\times
N\times N$.

Now we have fully classified all the truly entangled classes of
$2\times N\times N$ state. A truly entangled state of $2\times
N\times N$ must line in one of the sets $C_{N-i,l}$ (or $C_{n,l}$).
According to (\ref{nl-prop}) we can obtain a restriction on the
values of n,
\begin{eqnarray}
\frac{2N}{3}\leq & n & < N  \label{ineqn}
\end{eqnarray}
when $i >0$, and from the arguments above (\ref{jandjprime}) we know
\begin{eqnarray}
n =  N \; ,\;  0 <  l  < N \label{ineql}
\end{eqnarray}
when $i=0$. From those two theorems proved in this section we know
that the mapping of $C_{N-i,l} \mapsto c_{N-i,l}$ determines all the
true entanglement classes in $C_{N-i,l}$.

\section{Examples}

According to above explanation, the classification of the entangled
state $2\times N \times N$ may be accomplished by repeatedly taking
the above introduced procedures. To be more specific and for readers
convenience, in the following we completely classify the $2\times
2\times 2$ and $2\times 4\times 4$ pure states by using this novel
method as examples.

For $N=2$, i.e., three qubits states, there is only one inequivalent
set $c_{N=2,\,l=1}$ (the case $c_{1,1}$ does not exist in the three
qubits true entanglement state due to proposition (\ref{nl-prop})).
There are two inequivalent Jordan forms for $2\times 2$ matrices
with rank one, and thus two inequivalent classes in $c_{2,1}$ which
correspond to the GHZ and W states separately \cite{three-qubit}
\begin{eqnarray}
\mathrm{GHZ} &:& \; E=
\left[
\begin{array}{ll}
1 & 0 \\
0 & 1
\end{array}
\right],\; J = \left[
\begin{array}{ll}
\lambda & 0 \\
0 & 0
\end{array}
\right]  , \\
\mathrm{W} &:& \; E=
\left[
\begin{array}{ll}
1 & 0 \\
0 & 1
\end{array}
\right],\; J= \left[
\begin{array}{ll}
0 & 1 \\
0 & 0
\end{array}
\right] .
\end{eqnarray}

For $N=4$, from (\ref{nl-prop})-(\ref{ineql}) the inequivalent sets
are
\begin{eqnarray}
&&c_{N,\,l}  =  \, c_{4,1}\, , c_{4,2}\, , c_{4,3}  \\
&&c_{N-1,\,l}  =  \, c_{3,2}\, , c_{3,3}\; .
\end{eqnarray}
Due to (\ref{ineqn}), there is no $c_{N-i,\,l}$ sets in truly
entangled states for $N=4$ when $i\geq 2$. In the case of
$c_{4,\,l}$ all
inequivalent classes have the form of $\left(\begin{array}{l} E \\
J \end{array}\right)$, where
\begin{eqnarray}
E = \left[
  \begin{array}{cccccc}
    1 & 0 & 0 & 0 \\
    0 & 1 & 0 & 0 \\
    0 & 0 & 1 & 0 \\
    0 & 0 & 0 & 1 \\
  \end{array}
\right] .
\end{eqnarray}
Hence, we can distinguish them just by virtue of $J$'s pattern.
There are two classes in set $c_{4,1}$, i.e.,
\begin{eqnarray}
\left[
  \begin{array}{cccccc}
    \lambda & 0 & 0 & 0 \\
    0 & 0 & 0 & 0 \\
    0 & 0 & 0 & 0 \\
    0 & 0 & 0 & 0 \\
  \end{array}
\right]\;, \left[
  \begin{array}{cccccc}
    0 & 1 & 0 & 0 \\
    0 & 0 & 0 & 0 \\
    0 & 0 & 0 & 0 \\
    0 & 0 & 0 & 0 \\
  \end{array}
\right]\; ;
\end{eqnarray}
six classes in set $c_{4,2}$, the
\begin{eqnarray}
\left[
  \begin{array}{cccccc}
    \lambda_{1} & 0 & 0 & 0 \\
    0 & \lambda_{2} & 0 & 0 \\
    0 & 0 & 0 & 0 \\
    0 & 0 & 0 & 0 \\
  \end{array}
\right]\; , \left[
\begin{array}{cccccc}
    \lambda & 0 & 0 & 0 \\
    0 & 0 & 1 & 0  \\
    0 & 0 & 0 & 0  \\
    0 & 0 & 0 & 0  \\
\end{array}
\right]\;, \left[
\begin{array}{cccccc}
    0 & 1 & 0 & 0 \\
    0 & 0 & 1 & 0 \\
    0 & 0 & 0 & 0 \\
    0 & 0 & 0 & 0 \\
\end{array}
\right]\; , \nonumber\\
\left[
  \begin{array}{cccccc}
    0 & 1 & 0 & 0 \\
    0 & 0 & 0 & 0 \\
    0 & 0 & 0 & 1 \\
    0 & 0 & 0 & 0 \\
  \end{array}
\right]\; , \left[
  \begin{array}{cccccc}
    \lambda & 0 & 0 & 0 \\
    0 & \lambda & 0 & 0 \\
    0 & 0 & 0 & 0  \\
    0 & 0 & 0 & 0  \\
  \end{array}
\right]\; , \left[
  \begin{array}{cccccc}
    \lambda & 1 & 0 & 0 \\
    0 & \lambda & 0 & 0 \\
    0 & 0 & 0 & 0 \\
    0 & 0 & 0 & 0 \\
  \end{array}
\right]\; \label{2time4times4} ;
\end{eqnarray}
and five classes in set $c_{4,3}$, the
\begin{eqnarray}
 \left[
  \begin{array}{cccccc}
    \lambda_{1} & 0 & 0 & 0 \\
    0 & \lambda_{2} & 0 & 0 \\
    0 & 0 & \lambda_{3} & 0 \\
    0 & 0 & 0 & 0 \\
  \end{array}
\right]\; , \left[
  \begin{array}{cccccc}
    \lambda_{1} & 0 & 0 & 0 \\
    0 & \lambda_{2} & 0 & 0 \\
    0 & 0 & 0 & 1 \\
    0 & 0 & 0 & 0 \\
  \end{array}
\right]\; , \left[
  \begin{array}{cccccc}
    \lambda & 0 & 0 & 0 \\
    0 & 0 & 1 & 0 \\
    0 & 0 & 0 & 1 \\
    0 & 0 & 0 & 0 \\
  \end{array}
\right]\; , \left[
  \begin{array}{cccccc}
    0 & 1 & 0 & 0 \\
    0 & 0 & 1 & 0 \\
    0 & 0 & 0 & 1 \\
    0 & 0 & 0 & 0 \\
  \end{array}
\right]\; , \left[
  \begin{array}{cccccc}
    \lambda & 1 & 0 & 0 \\
    0 & \lambda & 0 & 0 \\
    0 & 0 & 0 & 1 \\
    0 & 0 & 0 & 0 \\
  \end{array}
\right]\; .\;~~~
\end{eqnarray}

In the case of $c_{3,\,l}$, every class has the form of
$\left(\begin{array}{l} \Lambda \\ \Gamma \end{array}\right)$ where
\begin{eqnarray}
\Lambda = \left[
  \begin{array}{cccccc}
    1 & 0 & 0 & 0 \\
    0 & 1 & 0 & 0 \\
    0 & 0 & 1 & 0 \\
    0 & 0 & 0 & 0 \\
  \end{array}
\right]\; .
\end{eqnarray}
From theorem 2 in Section \ref{section-theorem2} we can simply
classify the set $c_{3,2}$ by the pattern of $\Gamma$ matrix. And,
from the measure in constructing matrices $B_3$ and $B_4$ in the
same section, we find that there is one class in $c_{3,2}$
\begin{eqnarray}
\left[
  \begin{array}{cccccc}
    0 & 0 & 0 & 0 \\
    0 & 0 & 0 & 0 \\
    0 & 0 & 0 & 1 \\
    0 & 1 & 0 & 0 \\
  \end{array}
\right]\; ,
\end{eqnarray}
and two classes in $c_{3,3}$
\begin{eqnarray}
\left[
  \begin{array}{cccccc}
    0 & 0 & 1 & 0 \\
    0 & 0 & 0 & 0 \\
    0 & 0 & 0 & 1 \\
    0 & 1 & 0 & 0 \\
  \end{array}
\right]\; , \; \left[
  \begin{array}{cccccc}
    0 & 0 & 0 & 0 \\
    1 & 0 & 0 & 0 \\
    0 & 0 & 0 & 1 \\
    0 & 1 & 0 & 0 \\
  \end{array}
\right] . \label{symmetry}
\end{eqnarray}
Altogether, there are 16 genuine entanglement classes in $2\times
4\times 4$ states, which agrees with what obtained in
Ref.\cite{range-2}. From (\ref{symmetry}) one can easily conclude
that the permutation of the two $4$ dimension partites are sorted
into different classes, which was noticed in \cite{range-2}.

\begin{figure}[t,m,u]
\centering
\includegraphics[width=5cm,height=5cm]{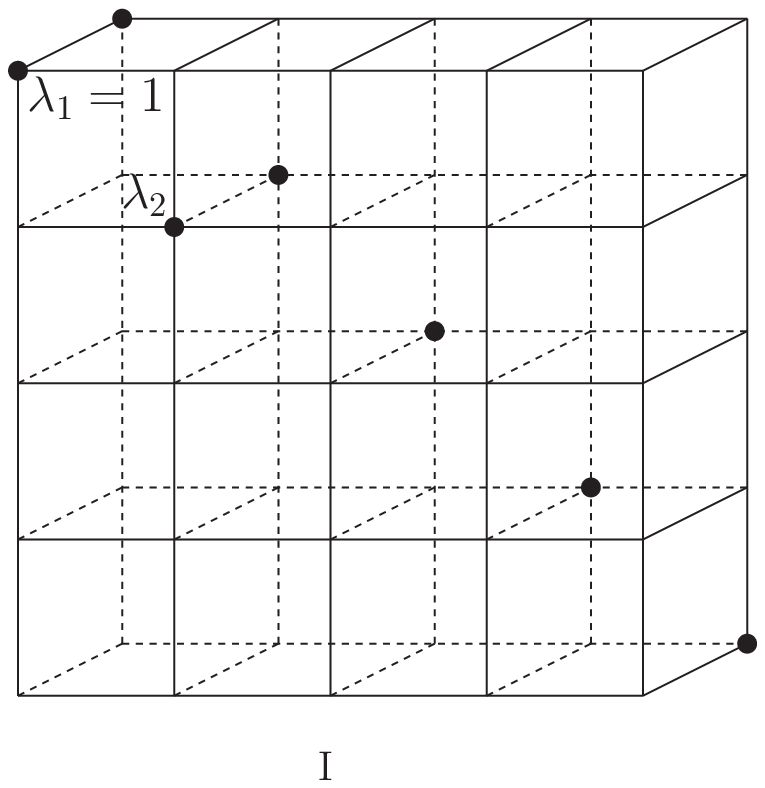}\;%
\includegraphics[width=5cm,height=5cm]{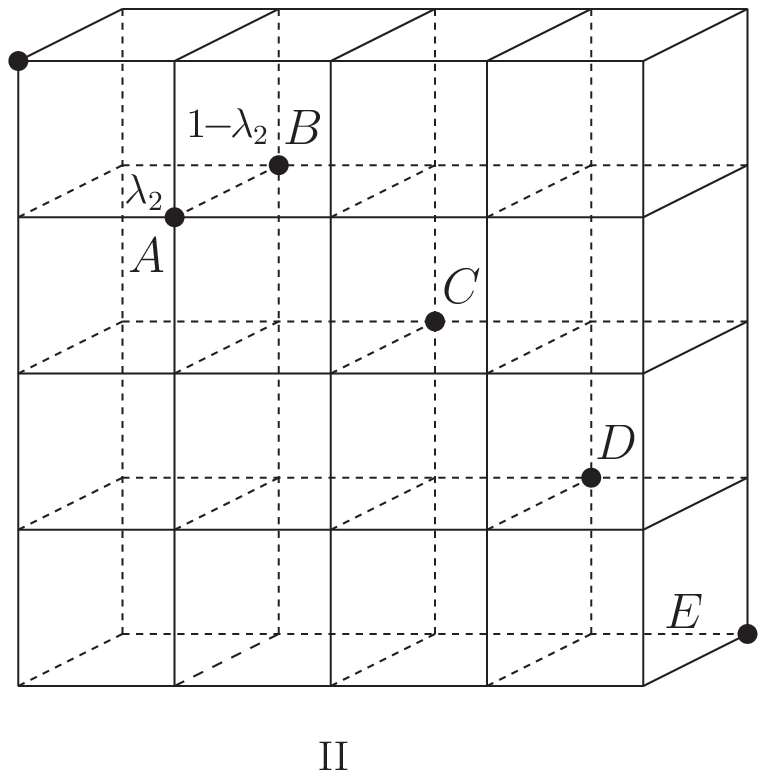}\\%
\includegraphics[width=5cm,height=5cm]{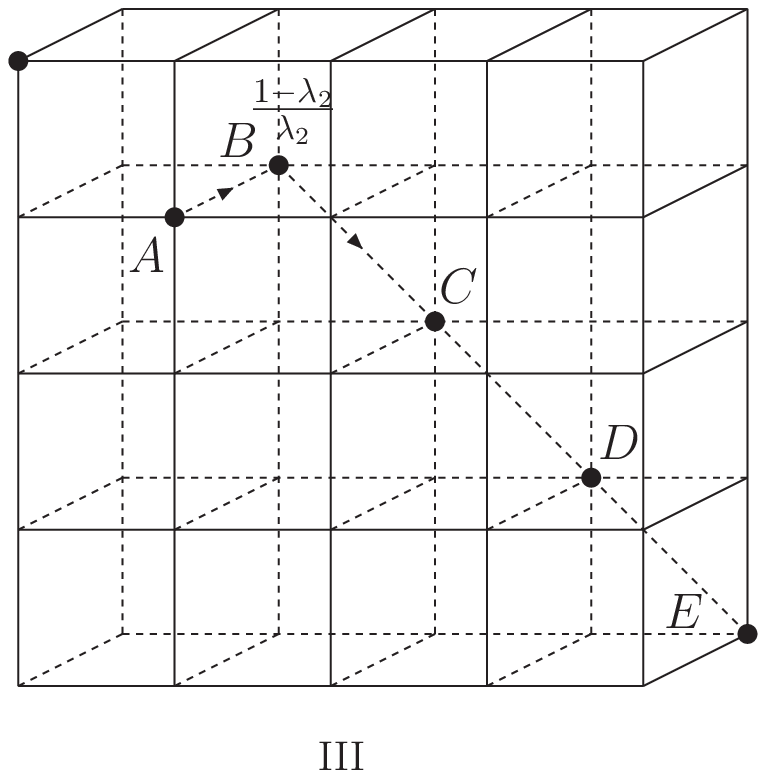}%
\includegraphics[width=5cm,height=5cm]{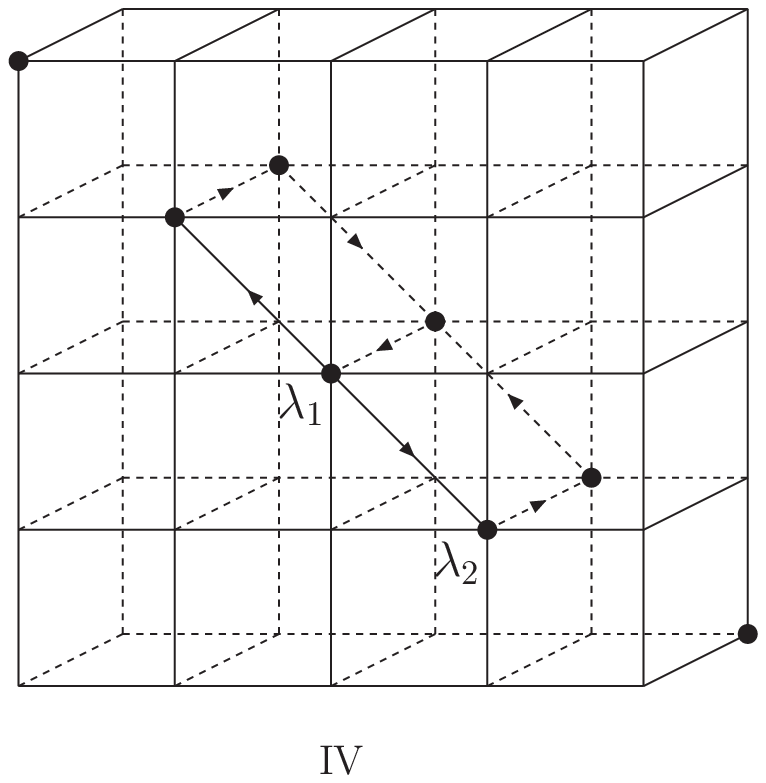}
\caption{\small The pictorial procedure of eliminating the
parameters in entanglement state. The plain nodes represent 0 and
the solid dark nodes represent 1 if not further specified. (I)
represents the initial state in cubic form; (II) is a transformed
form from (I) by subtracting the front plane from the back one;
(III) shows the elimination procedure of the parameter on node A.
(IV) shows the case with two free parameters. }\label{Pic-f-5-3}
\end{figure}

In above examples we enumerate various distinct classes of the
$2\times 2\times 2$ and $2\times 4\times 4$ states, each with a
representative state. In fact there are still reducible parameters
in the representative states, the eigenvalues of the Jordan form,
e.g. the $\lambda$s in (\ref{2time4times4}). These parameters can be
sorted into two categories: one with only redundant parameters,
which can be eliminated out of the state through ILOs; another
possesses non-local parameters which can not be eliminated through
the ILOs and will keep on staying in the entangled state as free
parameters. For the first case, we take one typical class in set
$c_{5,\,2}$ as an example. The first three diagrams of
Fig.(\ref{Pic-f-5-3}) exhibits the procedure of how the redundant
parameters being eliminated through elementary operations.
Multiplying the vertical or horizontal A-B plane of the cubic form
in (II) by a factor of $\frac{1}{\lambda_{2}}$ (elementary operation
{\bf type 2}), we can get the form of diagram (III). The
multiplication of the back plane of the cubic by factor of
$\frac{\lambda_{2}}{1-\lambda_{2}}$ (elementary operation {\bf type
2}) will transform the parameter $\frac{1-\lambda_{2}}{\lambda_{2}}$
from node B to nodes C, D, E, which is represented by the arrow
between B and C. Then, the parameters can be easily eliminated by
elementary operations in three vertical planes containing nodes C,
D, and E, respectively.

For the second case, different from the situations shown in first
three diagrams, the flow of parameters in performing the elementary
operations may form loops, like shown in diagram IV of
Fig.(\ref{Pic-f-5-3}) for one typical class in set $c_{5,\,4}$ as an
example. As long as the loop(s) appears, the non-local parameter(s)
in the entanglement state remains, and vice versa. The number of
non-local parameters therefore equals to the number of the loops. It
is worth to mention that although these parameters are free ones,
they satisfy certain relations in giving out the equivalent classes,
like
\begin{eqnarray}
\left\lgroup \begin{array}{c} \lambda_{1}\\ \lambda_{2} \\
\end{array} \right\rgroup \sim \left\lgroup \begin{array}{c} \lambda_{2}\\
\lambda_{1} \\
\end{array} \right\rgroup \sim \left\lgroup \begin{array}{c} 1-\lambda_{1} \\
1-\lambda_{2}
\\ \end{array} \right\rgroup \sim \left\lgroup \begin{array}{c}
\frac{1}{\lambda_{1}} \\ \frac{1}{\lambda_{2}} \\ \end{array}
\right\rgroup \sim \left\lgroup \begin{array}{c} \frac{1}{\lambda_{1}}\\
\frac{\lambda_{2}}{\lambda_1}
\\ \end{array} \right\rgroup
\end{eqnarray}
for $c_{5,\,4}$. The situation may become complicated as the number
of parameters increase. To get a deeper insight of the behavior of
those non-local parameters, there still needs a lot of work.

\section{Summary and Conclusions}

In conclusion, we put forward a novel method in classifying the
entangled pure states of $2\times N\times N$. A remarkable feature
of our method in different from what existed in the literature is
that it does not need to classify the lower dimension cases first
\cite{range-1,range-2}. We find in practice that this method in
classifying the $2\times N\times N$ tri-partite entanglement state
is quite straightforward. Since the software for Jordan
decomposition is available, this new method can be applied to the
classification of a given state via automatic computer calculation,
which is very important as the partite dimension $N$ tends to be
large. Last, but not least, in this work a pictorial configuration
of the entanglement states on the grids is proposed , which gives an
intuitive demonstration for the non-local parameters, and is
efficient in eliminating redundant parameters.
\vspace{1.3cm}
\par
\par
{\bf Acknowledgments} \vspace{.2cm}

This work was supported in part by the National Natural Science
Foundation of China(NSFC) with numbers 10491306,10521003,10775179
and by the Scientific Research Fund of GUCAS with number 055101BM03.
C.-F.Q. is grateful to the hospitality of Kavli Institute for
Theoretical Physics China for a visit, when part of this work was
done.
\newpage
\appendix{\bf\Large Appendix}

\section{The Construction of B matrix} \label{construct-B-matrix}

\begin{eqnarray}
 P_1 \Gamma_{\!1}' Q_1 & = &
\left(
\begin{array}{llllll}
1 & 0 & 0 & 0 & 0 & 0 \\
0 & 1 & 0 & 0 & 0 & 0 \\
0 & 0 & 1 & 0 & 0 & 0 \\
0 & 0 & 0 & 1 & 0 & 0 \\
0 & 0 & 0 & 0 & 1 & 0 \\
0 & 0 & 0 & 0 & 0 & 0
\end{array}
\right) \label{550}\\ [0.2cm] P_1 \Gamma_{\!\!2}' Q_1 & = & \left(
\begin{array}{llllll}
\gamma_{11} & \gamma_{12} & \gamma_{13} & \gamma_{14} & \gamma_{15}
& \gamma_{16} \\ \gamma_{21} & \gamma_{22} & \gamma_{23} &
\gamma_{24} & \gamma_{25} & \gamma_{26}  \\ \gamma_{31} &
\gamma_{32} & \gamma_{33} & \gamma_{34} & \gamma_{35} & \gamma_{36}
\\ \gamma_{41} & \gamma_{42} & \gamma_{43} & \gamma_{44} & \gamma_{45} &
\gamma_{46}  \\ \gamma_{51} & \gamma_{52} & \gamma_{53} &
\gamma_{54} & \gamma_{55} & \gamma_{56}  \\ \gamma_{61} &
\gamma_{62} & \gamma_{63} & \gamma_{64} & \gamma_{65} & \gamma_{66}
\end{array}
\right). \label{551}
\end{eqnarray}

A direct observation on Eq.(\ref{551}) tells that $\gamma_{66}$ must
be zero, otherwise one can find invertible operators $P_{x}$ and
$Q_{x}$ which enable
\begin{eqnarray}
 P_{x}P_1 \Gamma_{\!1}' Q_1 Q_{x} & = &
\left(
\begin{array}{llllll}
1 & 0 & 0 & 0 & 0 & 0 \\
0 & 1 & 0 & 0 & 0 & 0 \\
0 & 0 & 1 & 0 & 0 & 0 \\
0 & 0 & 0 & 1 & 0 & 0 \\
0 & 0 & 0 & 0 & 1 & 0 \\
0 & 0 & 0 & 0 & 0 & 0
\end{array}
\right) \; , \\ [0.2cm] P_{x}P_1 \Gamma_{\!\!2}' Q_1Q_{x} & = &
\left(
\begin{array}{llllll}
\gamma_{11x} & \gamma_{12x} & \gamma_{13x} & \gamma_{14x} &
\gamma_{15x} & 0 \\ \gamma_{21x} & \gamma_{22x} & \gamma_{23x} &
\gamma_{24x} & \gamma_{25x} & 0 \\ \gamma_{31x} & \gamma_{32x} &
\gamma_{33x} & \gamma_{34x} & \gamma_{35x} & 0 \\ \gamma_{41x} &
\gamma_{42x} & \gamma_{43x} & \gamma_{44x} & \gamma_{45x} & 0
\\ \gamma_{51x} & \gamma_{52x} & \gamma_{53x} & \gamma_{54x} &
\gamma_{55x} & 0  \\ 0 & 0 & 0 & 0 & 0 & 1
\end{array} \right)\; .
\label{gamma-55-0}
\end{eqnarray}
Given $\lambda_{i}$ are the eigenvalues of submatrix
$\{\gamma_{ijx}\}_{5\times 5}$ and $\lambda \neq -\lambda_{i}$, we
will find that $r(P_{x}P_1 \Gamma_{\!\!2}' Q_1Q_{x} + \lambda
P_{x}P_1 \Gamma_{\!1}' Q_1 Q_{x}) = N > N-1$. This contradicts to
requirement that the maximum rank of $(t_{11} \Gamma_{\! 1} + t_{12}
\Gamma_{\!\!2})$ is $N-1$.

Let $\gamma_{66} = 0$, Eqs.(\ref{550}) and (\ref{551}) become
\begin{eqnarray}
 P_1 \Gamma_{\!1}' Q_1 & = &
\left(
\begin{array}{llllll}
1 & 0 & 0 & 0 & 0 & 0 \\
0 & 1 & 0 & 0 & 0 & 0 \\
0 & 0 & 1 & 0 & 0 & 0 \\
0 & 0 & 0 & 1 & 0 & 0 \\
0 & 0 & 0 & 0 & 1 & 0 \\
0 & 0 & 0 & 0 & 0 & 0
\end{array}
\right)\; , \\ [0.2cm] P_1 \Gamma_{\!\!2}' Q_1 & = & \left(
\begin{array}{llllll}
\gamma_{11} & \gamma_{12} & \gamma_{13} & \gamma_{14} & \gamma_{15}
& \gamma_{16} \\ \gamma_{21} & \gamma_{22} & \gamma_{23} &
\gamma_{24} & \gamma_{25} & \gamma_{26}  \\ \gamma_{31} &
\gamma_{32} & \gamma_{33} & \gamma_{34} & \gamma_{35} & \gamma_{36}
\\ \gamma_{41} & \gamma_{42} & \gamma_{43} & \gamma_{44} &
\gamma_{45} & \gamma_{46}  \\ \gamma_{51} & \gamma_{52} &
\gamma_{53} & \gamma_{54} & \gamma_{55} & \gamma_{56}  \\
\gamma_{61} & \gamma_{62} & \gamma_{63} & \gamma_{64} & \gamma_{65}
& 0
\end{array}
\right)\; .\label{gamma}
\end{eqnarray}
Since we are considering the true entanglement of $2\times N\times
N$ states, neither the last column nor the last row of the matrix in
Eq.(\ref{gamma}) can be completely zero. There exist ILOs $P_{2},
Q_{2}$ which satisfy the following equations
\begin{eqnarray}
P_{2}P_1 \Gamma_{\!1}' Q_1Q_{2} & = & \left(
\begin{array}{llllll}
1 & 0 & 0 & 0 & 0 & 0 \\
0 & 1 & 0 & 0 & 0 & 0 \\
0 & 0 & 1 & 0 & 0 & 0 \\
0 & 0 & 0 & 1 & 0 & 0 \\
0 & 0 & 0 & 0 & 1 & 0 \\
0 & 0 & 0 & 0 & 0 & 0
\end{array}
\right)\; , \\ [0.2cm] P_{2}P_1 \Gamma_{\!\!2}' Q_1Q_{2} & = &
\left(
\begin{array}{llllll}
\gamma_{11}' & \gamma_{12}' & \gamma_{13}' & \gamma_{14}' &
\gamma_{15}' & 0 \\ \gamma_{21}' & \gamma_{22}' & \gamma_{23}' &
\gamma_{24}' & \gamma_{25}' & 0 \\ \gamma_{31}' & \gamma_{32}' &
\gamma_{33}' & \gamma_{34}' & \gamma_{35}' & 0 \\ \gamma_{41}' &
\gamma_{42}' & \gamma_{43}' & \gamma_{44}' & \gamma_{45}' & 0  \\
\gamma_{51}' & \gamma_{52}' & \gamma_{53}' & \gamma_{54}' &
\gamma_{55}' & 1  \\ \gamma_{61}' & \gamma_{62}' & \gamma_{63}' &
\gamma_{64}' & \gamma_{65}' & 0
\end{array}
\right)\; .
\end{eqnarray}
An invertible operator $Q_{3}$
\begin{eqnarray}
Q_{3} = \left(
\begin{array}{llllll}
1 & 0 & 0 & 0 & 0 & 0 \\
0 & 1 & 0 & 0 & 0 & 0  \\
0 & 0 & 1 & 0 & 0 & 0  \\
0 & 0 & 0 & 1 & 0 & 0  \\
0 & 0 & 0 & 0 & 1 & 0  \\
-\gamma_{51}' & -\gamma_{52}' & -\gamma_{53}' & -\gamma_{54}' &
-\gamma_{55}' & 1
\end{array}
\right)
\end{eqnarray}
makes
\begin{eqnarray}
P_{2}P_1 \Gamma_{\!1}' Q_1Q_{2}Q_{3} & = & \left(
\begin{array}{llllll}
1 & 0 & 0 & 0 & 0 & 0 \\
0 & 1 & 0 & 0 & 0 & 0 \\
0 & 0 & 1 & 0 & 0 & 0 \\
0 & 0 & 0 & 1 & 0 & 0 \\
0 & 0 & 0 & 0 & 1 & 0 \\
0 & 0 & 0 & 0 & 0 & 0
\end{array}
\right)\; , \label{gamma11} \\ [0.2cm] P_{2}P_1 \Gamma_{\!\!2}'
Q_1Q_{2}Q_{3} & = & \left(
\begin{array}{llllll}
\gamma_{11}' & \gamma_{12}' & \gamma_{13}' & \gamma_{14}' &
\gamma_{15}' & 0 \\ \gamma_{21}' & \gamma_{22}' & \gamma_{23}' &
\gamma_{24}' & \gamma_{25}' & 0 \\ \gamma_{31}' & \gamma_{32}' &
\gamma_{33}' & \gamma_{34}' & \gamma_{35}' & 0 \\ \gamma_{41}' &
\gamma_{42}' & \gamma_{43}' & \gamma_{44}' & \gamma_{45}' & 0  \\
0 & 0 & 0 & 0 & 0 & 1
\\ \gamma_{61}' & \gamma_{62}' & \gamma_{63}' & \gamma_{64}' &
\gamma_{65}' & 0
\end{array}
\right)\; .\label{gamma54}
\end{eqnarray}
Here, $\gamma_{65}' $ must be zero also, otherwise to keep the form
of (\ref{gamma11}) unchanged the matrix in Eq.(\ref{gamma54}) can be
transformed into
\begin{eqnarray}
\left(
\begin{array}{llllll}
\gamma_{11}'' & \gamma_{12}'' & \gamma_{13}'' & \gamma_{14}'' & 0 &
0 \\ \gamma_{21}'' & \gamma_{22}'' & \gamma_{23}'' & \gamma_{24}'' &
0 & 0 \\ \gamma_{31}'' & \gamma_{32}'' & \gamma_{33}'' &
\gamma_{34}'' & 0 & 0 \\ \gamma_{41}'' & \gamma_{42}'' &
\gamma_{43}'' & \gamma_{44}'' & 0 & 0  \\ 0 & 0 & 0 & 0 & 0 & 1  \\
0 & 0 & 0 & 0 & 1 & 0
\end{array}
\right)\; .
\end{eqnarray}
Clearly this will lead to the same contradiction as $\gamma_{66}$
does in (\ref{551}) and (\ref{gamma-55-0}). Thus Eq.(\ref{gamma54})
becomes
\begin{eqnarray}
P_{2}P_1 \Gamma_{\!\!2}' Q_1Q_{2}Q_{3} & = & \left(
\begin{array}{llllll}
\gamma_{11}' & \gamma_{12}' & \gamma_{13}' & \gamma_{14}' &
\gamma_{15}' & 0 \\ \gamma_{21}' & \gamma_{22}' & \gamma_{23}' &
\gamma_{24}' & \gamma_{25}' & 0 \\ \gamma_{31}' & \gamma_{32}' &
\gamma_{33}' & \gamma_{34}' & \gamma_{35}' & 0 \\ \gamma_{41}' &
\gamma_{42}' & \gamma_{43}' & \gamma_{44}' & \gamma_{45}' & 0  \\
0 & 0 & 0 & 0 & 0 & 1
\\ \gamma_{61}' & \gamma_{62}' & \gamma_{63}' & \gamma_{64}' &
0 & 0
\end{array}
\right)\; .
\end{eqnarray}
Applying the same procedure to the last row as we have performed to
the last column, we have
\begin{eqnarray}
\Lambda = P \Gamma_{\!1}' Q & = & \left(
\begin{array}{llllll}
1 & 0 & 0 & 0 & 0 & 0 \\
0 & 1 & 0 & 0 & 0 & 0 \\
0 & 0 & 1 & 0 & 0 & 0 \\
0 & 0 & 0 & 1 & 0 & 0 \\
0 & 0 & 0 & 0 & 1 & 0 \\
0 & 0 & 0 & 0 & 0 & 0
\end{array}
\right)\; , \\
\Gamma_{\!2}'' = P \Gamma_{\!2}' Q & = & \left(
  \begin{array}{ccc|ccc}
    \times & \times & \times & 0 & c_{15} & 0 \\
    \times & \times & \times & 0 & c_{25} & 0 \\
    \times & \times & \times & 0 & c_{35} & 0 \\ \hline
    r_{31} & r_{32} & r_{33} & 0 & 0 & 0 \\
    0 & 0 & 0 & 0 & 0 & 1 \\
    0 & 0 & 0 & 1 & 0 & 0 \\
  \end{array}
\right) = \left(
\begin{array}{cc}
A & c \\
r & B_{3} \\
\end{array}
\right)\; ,
\end{eqnarray}
where $P=\prod_{i}P_{i}$ and $Q = \prod_i Q_{i}$ are sequences of
invertible operators $P_{i}$ and $Q_{i}$, respectively.

\section{The Superpositions of $\Lambda'$ and $B_n$}
\label{superposition-lambda-gamma}

Eq.(\ref{app-prove}) can be written in the following matrix
equations
\begin{eqnarray}
\left\{ \begin{array}{ccl}  P_{B_n}( \Lambda'_n + \lambda B_n
)Q_{B_n} & = & \Lambda'_n  \\ \\ P_{B_n} B_n Q_{B_n} & = & B_n
\end{array} \right. \; . \label{pbn-qbn}
\end{eqnarray}
Here, for the sake of clarity, we label the $\Lambda'$ with
subscript $n$ to indicate its dimension.

Following, we show inductively that the invertible matrices
$P_{B_n}$ and $Q_{B_n}$ can always be constructed.

First, in case $n=1$, then $\Lambda'=0$, $B_{1} = 0$, the
construction of invertible operators $P_{B_1}$, $Q_{B_1}$
Eq.(\ref{pbn-qbn}) is trivial.

In the case of $n=2$, Eq.(\ref{pbn-qbn}) becomes
\begin{eqnarray}
\left\{ \begin{array}{ccl}  P_{B_2} \left(
\begin{array}{cc}
1 & \lambda \\
0 & 0 \\
\end{array}
\right) Q_{B_2} & = & \left(
\begin{array}{cc}
1 & 0 \\
0 & 0 \\
\end{array}
\right)  \\ \\
P_{B_2} \left(
\begin{array}{cc}
0 & 1 \\
0 & 0 \\
\end{array}
\right) Q_{B_2} & = & \left(
\begin{array}{cc}
0 & 1 \\
0 & 0 \\
\end{array}
\right)
\end{array} \right. \; .
\end{eqnarray}
$P_{B_2} =E$ and $Q_{B_2}$ of the form $\left(
\begin{array}{cc}
1 & -\lambda \\
0 & 1 \\
\end{array}
\right)$ satisfy the above equations.

Suppose for arbitrary $n$ (\ref{pbn-qbn}) is true, we show that
$P_{B_{n+1}}, Q_{B_{n+1}}$ can also be constructed, satisfying
\begin{eqnarray}
\left\{ \begin{array}{ccl}  P_{B_{n+1}}( \Lambda'_{n+1} + \lambda
B_{n+1} )Q_{B_{n+1}} & = & \Lambda'_{n+1}  \\ \\ P_{B_{n+1}} B_{n+1}
Q_{B_{n+1}} & = & B_{n+1}
\end{array} \right. \; . \label{pbn+1}
\end{eqnarray}
Here, either
\begin{eqnarray}
B_{n+1} = \left(
  \begin{array}{cc}
    0 & r \\
    0 & B_{n} \\
  \end{array}
\right) \;\label{Bn+1a}
\end{eqnarray}
or
\begin{eqnarray}
 \; B_{n+1} = \left(
  \begin{array}{cc}
    0 & 0 \\
    c & B_{n} \\
  \end{array}
\right) \label{Bn+1b} \; ,
\end{eqnarray}
where $r(B_{n})=n-1$, the ranks of $\left(
\begin{array}{c}
r \\
 B_{n} \\
 \end{array}
 \right)$ and
$\left(
\begin{array}{cc}
c & B_{n} \\
\end{array}
\right)$ are $n$. And,
\begin{eqnarray}
\Lambda'_{n+1} = \left(
  \begin{array}{cc}
    1 & 0 \\
    0 & \Lambda'_{n} \\
  \end{array}
\right) \; .
\end{eqnarray}

In one example of $n=5$, $\Lambda'$ and $B$ can be expressed as
follows
\begin{eqnarray} \Lambda'_{5+1} = \left(
  \begin{array}{cccccc}
    1 & 0 & 0 & 0 & 0 & 0 \\
    0 & 1 & 0 & 0 & 0 & 0 \\
    0 & 0 & 1 & 0 & 0 & 0 \\
    0 & 0 & 0 & 1 & 0 & 0 \\
    0 & 0 & 0 & 0 & 1 & 0 \\
    0 & 0 & 0 & 0 & 0 & 0 \\
  \end{array}
\right),\;\; B_{5+1} = \left(
  \begin{array}{cccccc}
    0 & 0 & 1 & 0 & 0 & 0 \\
    0 & 0 & 0 & 0 & 0 & 0 \\
    0 & 0 & 0 & 0 & 1 & 0 \\
    0 & 1 & 0 & 0 & 0 & 0 \\
    0 & 0 & 0 & 0 & 0 & 1 \\
    0 & 0 & 0 & 1 & 0 & 0 \\
  \end{array}
\right)\; ,
\end{eqnarray}
where $B_{5+1} = \left(
  \begin{array}{cc}
    0 & r \\
    0 & B_{5} \\
  \end{array}
\right)$, $r=(0,1,0,0,0)$.

The operator $P_{B_{n+1}}$ and $Q_{B_{n+1}}$ can be constructed as
follows
\begin{eqnarray}
P_{B_{n+1}} & = & \left(
\begin{array}{cc}
1 & X \\
0 & E \\
\end{array}
\right) \left(
\begin{array}{cc}
1 & 0 \\
0 & P_{B_{n}} \\
\end{array}
\right) \; , \label{PBn+1} \\
Q_{B_{n+1}} & = & \left(
\begin{array}{cc}
1 & 0 \\
0 & Q_{B_{n}} \\
\end{array}
\right) \left(
\begin{array}{cc}
1 & -Y \\
0 & E \\
\end{array}
\right) \; , \label{QBn+1}
\end{eqnarray}
where $Y= \lambda r Q_{B_n} + X\Lambda'_{n}$. Because the rank of
$\left(
  \begin{array}{cc}
    0 & r  \\
    0 &  B_{n} \\
  \end{array}
\right)$ is unchanged under the invertible transformation, we can
always find such invertible operator $\left(
  \begin{array}{cc}
    1 & X  \\
    0 &  E \\
  \end{array}
\right) $ which satisfies
\begin{eqnarray}
\left(
  \begin{array}{cc}
    1 & X  \\
    0 &  E \\
  \end{array}
\right) \left(
  \begin{array}{cc}
    0 & r Q_{B_n}  \\
    0 &  B_{n} \\
  \end{array}
\right) = \left(
  \begin{array}{cc}
    0 &  r  \\
    0 &  B_{n} \\
  \end{array}
\right) \; .
\end{eqnarray}
It is then easy to verify Eq.(\ref{pbn+1}) using
Eqs.(\ref{PBn+1},\ref{QBn+1}). Note that the $P_{B_{n+1}}$ and
$Q_{B_{n+1}}$ constructed above correspond to the case of
Eq.(\ref{Bn+1a}), for the case of (\ref{Bn+1b}) the procedure is
similar.

Along the same line, it can also be found that there exist such
invertible operators $P'_{B_n}, Q'_{B_{n}}$ that
\begin{eqnarray}
\left\{ \begin{array}{ccl}  P'_{B_n} \Lambda'_n Q'_{B_n} & = &
\Lambda'_n  \\ \\ P'_{B_n} ( B_n + \lambda \Lambda'_{n} ) Q'_{B_n} &
= & B_n
\end{array} \right. \; . \label{inverse-pbn-qbn}
\end{eqnarray}

\section{The flip of $\Lambda'_n$ and $B_n$}
\label{flip-L-B}

Using the Eq.(\ref{pbn-qbn}) and Eq.(\ref{inverse-pbn-qbn}), we show
that there exist the following invertible matrices $P_t,\; Q_t$
which flip the $\Lambda'_n$ and $B_n$, like
\begin{eqnarray}
\left(
  \begin{array}{c}
    P_t \Lambda'_n Q_t \\
    P_t B_n Q_t \\
  \end{array}
\right) = \left(
  \begin{array}{c}
      B_n \\
     \Lambda'  \\
  \end{array}
\right)\label{flip-lamda-b} \; .
\end{eqnarray}

Provided
\begin{eqnarray}
\left\{ \begin{array}{ccl}  P_{B_n}(\lambda)( \Lambda'_n + \lambda
B_n )Q_{B_n}(\lambda) & = & \Lambda'_n  \\ \\ P_{B_n}(\lambda) B_n
Q_{B_n}(\lambda) & = & B_n
\end{array} \right. \; ,
\end{eqnarray}
and
\begin{eqnarray}
\left\{ \begin{array}{ccl}  P'_{B_n}(\lambda) \Lambda'_n
Q'_{B_n}(\lambda) & = & \Lambda'_n  \\ \\ P'_{B_n}(\lambda) ( B_n +
\lambda \Lambda'_{n} ) Q'_{B_n}(\lambda) & = & B_n
\end{array} \right. \; ,
\end{eqnarray}
it can be found that
\begin{eqnarray}
P_t  =  P_{B_n}(\lambda) P'_{B_n}(-\frac{1}{\lambda})
P_{B_{n}}(\lambda) \; ,\;\; Q_t  =  Q_{B_n}(\lambda)
Q'_{B_n}(-\frac{1}{\lambda}) Q_{B_{n}}(\lambda) \;
\end{eqnarray}
enables
\begin{eqnarray}
\left(
  \begin{array}{c}
    P_t \Lambda'_n Q_t \\
    P_t B_n Q_t \\
  \end{array}
\right) = \left(
  \begin{array}{c}
      -\lambda B_n \\
     \frac{1}{\lambda}\Lambda'_n  \\
  \end{array}
\right)\; ,
\end{eqnarray}
which is equivalent to (\ref{flip-lamda-b}) up to irrelevant
coefficients.

\end{document}